\documentclass[11pt]{article}

\usepackage{array}
\newcolumntype{P}[1]{>{\centering\arraybackslash}p{#1}}

\usepackage{booktabs} 

\usepackage{comment}

\usepackage{hyperref}

\usepackage{tikz}
\usepackage{verbatim}

\usepackage{amssymb}
\usepackage{lineno}
\usepackage{mathtools,bbm}
\usepackage{graphicx,url}
\usepackage{graphics}

\DeclareMathOperator*{\E}{\mathbb{E}}
\DeclareMathOperator*{\R}{\mathcal{R}}

\DeclareMathOperator*{\I}{\mathbbm{1}}
\DeclareMathOperator*{\F}{\mathcal{F}}

\DeclareMathOperator*{\argmax}{argmax}
\DeclareMathOperator*{\argmin}{argmin}

\usepackage{setspace,graphicx,epstopdf,amsmath,amsfonts,amssymb,amsthm,versionPO}
\usepackage{marginnote,datetime,enumitem,rotating,fancyvrb,scalerel}
\usepackage{hyperref}

\excludeversion{notes}		
\includeversion{links}          

\iflinks{}{\hypersetup{draft=true}}

\ifnotes{%
\usepackage[margin=1in,paperwidth=10in,right=2.5in]{geometry}%
\usepackage[textwidth=1.4in,shadow,colorinlistoftodos]{todonotes}%
}{%
\usepackage[margin=1in]{geometry}%
\usepackage[disable]{todonotes}%
}

\makeatletter\let\chapter\@undefined\makeatother 

\renewcommand{\thefootnote}{\fnsymbol{footnote}}

\usepackage{indentfirst} 
\usepackage{jfe}          

\newtheorem{proposition}{Proposition}

\newtheorem{definition}{Definition}

\usepackage[
  backend=biber,
  style=numeric, 
]{biblatex}
\usepackage{float}
\usepackage[caption = false]{subfig}

\begin{document}

\setlist{noitemsep}  

\title{\textbf{Deep Neural Networks for Choice Analysis:\\ Extracting Complete Economic Information for Interpretation}}

\author{Shenhao Wang \\
  Qingyi Wang \\
  Jinhua Zhao \\
  \\
  Massachusetts Institute of Technology \\
  77 Mass Ave, Cambridge, Massachusetts, U.S. \\
  }

\date{September 2020\footnote{Please cite: Wang, Wang, and Zhao (2020). ``Deep neural networks for choice analysis: Extracting complete economic information for interpretation.'' Transportation research part C: emerging technologies 118: 102701.}}              

\singlespacing
\maketitle

\vspace{-.2in}
\begin{abstract}
\noindent
While deep neural networks (DNNs) have been increasingly applied to choice analysis showing high predictive power, it is unclear to what extent researchers can interpret economic information from DNNs. This paper demonstrates that DNNs can provide economic information as \textbf{complete} as classical discrete choice models (DCMs). The economic information includes choice predictions, choice probabilities, market shares, substitution patterns of alternatives, social welfare, probability derivatives, elasticities, marginal rates of substitution, and heterogeneous values of time. Unlike DCMs, DNNs can automatically learn utility functions and reveal behavioral patterns that are not prespecified by domain experts, particularly when the sample size is large. However, the economic information obtained from DNNs can be unreliable when the sample size is small, because of three challenges associated with the automatic learning capacity: high sensitivity to hyperparameters, model non-identification, and local irregularity. The first challenge is related to the statistical challenge of balancing approximation and estimation errors of DNNs, the second to the optimization challenge of identifying the global optimum in the DNN training, and the third to the robustness challenge of mitigating locally irregular patterns of estimated functions. To demonstrate the strength and challenges, we estimated the DNNs using a stated preference survey from Singapore and a revealed preference data from London, extracted the full list of economic information from the DNNs, and compared them with those from the DCMs. We found that the economic information either aggregated over trainings or population is more reliable than the disaggregate information of the individual observations or trainings, and that larger sample size, hyperparameter searching, model ensemble, and effective regularization can significantly improve the reliability of the economic information extracted from the DNNs. Future studies should investigate the requirement of sample size, better ensemble mechanisms, other regularizations and DNN architectures, better optimization algorithms, and robust DNN training methods to address DNNs’ three challenges to provide more reliable economic information for DNN-based choice models. 
\end{abstract}

\medskip


\thispagestyle{empty}
\clearpage

\onehalfspacing
\setcounter{footnote}{0}
\renewcommand{\thefootnote}{\arabic{footnote}}
\setcounter{page}{1}

\section{Introduction}
\label{s:1}
\noindent
Discrete choice models (DCMs) have been used to examine individual decision making for decades with wide applications to economics, marketing, and transportation \cite{Ben_Akiva1985,Train2009}. Recently, there is an emerging trend of using machine learning models, particularly deep neural networks (DNNs), to analyze individual decisions. DNNs have shown its extraordinary predictive power across various academic disciplines \cite{LeCun2015}, and in the transportation field, DNNs achieve higher prediction accuracy than DCMs in predicting travel mode choice, automobile ownership, route choice, and many other tasks \cite{Nijkamp1996,Rao1998,XieChi2003,Cantarella2005,Celikoglu2006,Kaewwichian2019}. However, the interpretability of DNNs is relatively understudied despite the recent progresses \cite{Ribeiro2016,Boshi_Velez2017,ZhouBolei2014}: it remains unclear how to obtain reliable economic information from the DNNs in the context of choice analysis. 

This study demonstrates that DNNs can provide economic information as \textit{complete} as the classical DCMs, including choice predictions, choice probabilities, market share, substitution patterns of alternatives, social welfare, probability derivatives, elasticities, marginal rates of substitution (MRS), and heterogeneous values of time (VOT). The full list of economic information can be computed by using either the estimated utility and choice probability functions or the input gradients of these functions in DNNs. The process of interpreting DNNs for economic information is different from that of interpreting classical DCMs. The DNN interpretation has to be based on the full \textit{function} of choice probabilities, rather than the \textit{individual parameters} as in classical DCMs. With thousands of individual parameters existing in DNNs, it proves meaningless and unnecessary to delve into individual parameters to extract economic information. We compare the DNNs to the multinomial logit (MNL) model by applying them to a stated preference dataset of travel mode choice in Singapore and a revealed preference dataset in London, showing the robustness of our approach to diverse contexts. This process of interpreting DNNs for economic information can be applied to any choice analysis scenario.

While DNNs can automatically reveal utility functions and behavioral patterns,  this power of automatic utility learning comes with three challenges: (1) high sensitivity to hyperparameters, (2) model non-identification, and (3) local irregularity. The first refers to the fact that the estimated DNNs are highly sensitive to the selection of hyperparameters that control the DNN complexity. The second refers to the fact that the optimization in the DNN training often identifies the local minima or saddle points rather than the global optimum, depending on the initialization of the DNN parameters. The third refers to the fact that DNNs have locally irregular patterns such as exploding gradients and the lack of monotonicity to the extent that certain choice behavior revealed by DNNs is not reasonable. The three challenges are embedded respectively in the statistical, optimization, and robustness discussions about DNNs. They become more severe when the sample size is small, but are somewhat mitigated when the sample size is large. While all three challenges create difficulties in interpreting DNN models for economic information, our empirical experiments show that simple random hyperparameter searching, common regularization methods, model ensemble, and information aggregation can partially mitigate these issues. 

This study makes the following contributions. This is the first study that systematically discusses the interpretation of DNNs for economic information in choice analysis, and shows that DNNs can provide economic information as complete as classical DCMs. At the same time, we point out the three challenges of interpreting DNNs for reliable economic information, as well as their theoretical roots. The challenges are different from those in the classical DCMs: DCMs do not even have the notion of hyperparameters, and model non-identification and local irregularity are typically not problems in the DCMs. While this study cannot fully address the challenges in DNN-based choice models, we demonstrate the importance of using large samples, hyperparameter searching, model ensemble, and regularization methods to improve the reliability of the economic information extracted from the DNNs. The paper provides a practical guide for transportation modelers and methodological benchmarks for future researchers to compare to and improve upon. For future researchers to replicate our work, we uploaded our codes to a Github repository: \url{https://github.com/cjsyzwsh/dnn-for-economic-information.git}.

The paper is structured as follows. Section 2 reviews the studies about DCMs and DNNs concerning prediction, interpretability, sensitivity to hyperparameters, model non-identification, and local irregularity. Section 3 introduces the theory, models, and methods of computing economic information. Section 4 sets up the experiments, and Section 5 discusses the list of economic information obtained from the DNNs. Section 6 concludes the study and discusses the limitations, challenges, and future research.

\section{Literature Review}
\label{s:2}
\noindent
DCMs have been used for decades to analyze the choice of travel modes, travel frequency, travel scheduling, destination and origin, travel route, activities, location, car ownership, and many other decisions in the transportation field \cite{Ben_Akiva1996,Cantarella2005,Ben_Akiva2014,Small2007td,Ortuzar2011,Guan2018}. While demand forecasting is important in these applications, all the economic information provides insights to guide policy interventions. For example, market shares can be computed from the DCMs to understand the market power of competing industries \cite{Train2009}. Elasticities of travel demand describe how effective it is to influence travel behavior through the change of tolls or subsidies \cite{Small1998,Helveston2015}. VOT, as one important instance of MRS, can be used to measure the monetary gain of saved time after the improvement of a transportation system in a benefit-cost analysis  \cite{Small1998,Small2007td}. 

Recently researchers started to use machine learning models to analyze individual decisions. Karlaftis and Vlahogianni (2011) \cite{Karlaftis2011} summarized $86$ studies in six transportation fields in which DNNs were applied. Researchers used DNNs to predict travel mode choice \cite{Cantarella2005}, car ownership \cite{Paredes2017}, travel accidents \cite{ZhangZhenhua2018}, travelers' decision rules \cite{Cranenburgh2019}, driving behaviors \cite{HuangXiuling2018}, trip distribution \cite{Mozolin2000}, hierarchical demand structure \cite{WuXin2018}, queue lengths \cite{LeeSeunghyeon2019}, parking occupancy \cite{YangShuguan2019}, metro passenger flows \cite{HaoSiyu2019}, and traffic flows \cite{Polson2017,LiuLijuan2017,WuYuankai2018,ZhangJunbo2018,DoLoan2019,MaTao2020}. DNNs are also used to complement the smartphone-based survey \cite{XiaoGuangnian2016}, improve survey efficiency \cite{Seo2017}, synthesize new population \cite{Borysov2019}, and impute survey data \cite{Duan2016}. In the studies that focus on prediction accuracy, researchers often compare many classifiers, including DNNs, support vector machines, decision trees, random forests, and DCMs, and typically find that DNNs and RF perform better than the classical DCMs \cite{Pulugurta2013,Omrani2015,Sekhar2016,Hagenauer2017,Cantarella2005}. In other fields, researchers also found the superior performance of DNNs in prediction compared to all the other machine learning (ML) classifiers \cite{Fernandez2014,Kotsiantis2007}. 

Since DNNs are often criticized as a ``black-box'' model, many resent studies have investigated how to improve its interpretability \cite{Boshi_Velez2017}. Researchers distilled knowledge from DNNs by re-training an interpretable model to fit the predicted soft labels of a DNN \cite{Hinton2015}, visualizing hidden layers in convolutional neural networks \cite{ZhouBolei2014,Zeiler2014}, using salience or attention maps to identify important inputs \cite{Lipton2016}, computing input gradients with sensitivity analysis \cite{Baehrens2010,Selvaraju2017,Smilkov2017,Erhan2009}, using instance-based methods to identify representative individuals for each class \cite{Aamodt1994,Erhan2009,Simonyan2013}, or locally approximating functions to make models more interpretable \cite{Ribeiro2016}. In the transportation field, only a very small number of studies touched upon the interpretability issue of DNNs for the choice analysis. For example, researchers extracted the elasticity values from DNNs \cite{Rao1998}, ranked the importance of DNN input variables \cite{Hagenauer2017}, or visualized the input-output relationship to improve the understanding of DNN models \cite{Bentz2000}. However, no study has discussed systematically how to compute all the economic information from DNNs, and none have demonstrated the specific practical and theoretical challenges in the process of interpreting DNNs for economic information. The challenges include at least three types, and they are specific to DNNs but not DCMs.

First, DNN performance is highly sensitive to the choice of hyperparameters and model complexity, which is essentially a statistical challenge of balancing approximation and estimation errors. Mathematically, model performance is evaluated by the excess error \footnote{This excess error can also be referred to as generalization error, since it measures the capacity of estimated $\hat{f}$ being generalized to other contexts.}, defined as ${\E}_S [L(\hat{f}) - L(f^*)]$, in which $L = {\E}_{x,y}[l(y, f(x))]$ and $l(y, f(x))$ is the loss function; $f^*$ is the true data generating function; $S$ is the sample $\{(x_i, y_i)_1^N\}$. This excess error can be further decomposed: ${\E}_S [L(\hat{f}) - L(f^*)] = {\E}_S [L(\hat{f}) - L(f_F) + L(f_F) - L(f^*)]$, in which ${\E}_S [L(\hat{f}) - L(f_F)]$ is referred to as the estimation error and ${\E}_S [L(f_F) - L(f^*)]$ as the approximation error; $f_F$ is the best function to approximate true model $f^*$ in the hypothesis function space $F$. A complex model tends to have larger estimation errors and smaller approximation errors, and a simple model is the opposite. DNNs have very small approximation errors because it has been proven to be a universal approximator \cite{Hornik1989,Hornik1991,Cybenko1989}, which also leads to the large estimation error as an issue. The large estimation error in DNNs can be examined by using statistical learning theory \cite{Bousquet2004,Von_Luxburg2011,Vapnik1999,Wainwright2019,Vershynin2018}. Formally, the model complexity can be measured by the Vapnik-Chervonenkis (VC) dimension ($v$), which provides an upper bound on DNNs' estimation error (proof is available in Appendix I). Recently, progress has been made to provide a tighter upper bound on the estimation error of DNNs by using other methods \cite{Bartlett2002,Anthony2009,Neyshabur2015,Golowich2017}. It is important to note that to select DNNs' hyperparameters is to control DNNs' model complexity, which balances between approximation and estimation errors. When either the approximation errors or the estimation errors are high, the overall DNN performance is low. 

Researchers can control the model complexity of DNNs by using architectural and regularization hyperparameters, hyperparameter optimization methods, and model ensemble. In a standard feedforward DNN, the architectural hyperparameters include depth and width, and the regularization hyperparameters include the $L_1$ and $L_2$ penalty constants, training iterations, minibatch sizes, data augmentation, dropouts, early stopping, and others \cite{Goodfellow2016,Bishop2006,Krizhevsky2012,Vincent2008,ZhangChiyuan2016}. To choose from the large number of hyperparameters, researchers used random search \cite{Bergstra2012,Bergstra2011}, grid search \cite{Bergstra2011,Bergstra2012}, Gaussian process \cite{Snoek2012}, multi-fidelity optimization \cite{Falkner2018,LiLisha2017}, or even reinforecement learning \cite{Zoph2016}. Besides the common regularization hyperparameters, model ensemble is a particularly useful way to reduce the excess error of DNNs. Hansen and Salamon (1990) formally introduced neural network ensemble to improve DNNs' generalizability \cite{Hansen1990}. Researchers proved that the generalization error of an ensemble model is always smaller than the average of individual models \cite{Krogh1995}. Recently, researchers used neural network ensembles to predict financial behavior and human activities, showing the superior power of model ensemble over individual models \cite{Tsai2008,Irvine2020}. In our study, while it is impossible to incorporate all these methods, we will demonstrate that several baseline methods are effective in improving the reliability of economic information in DNNs. 

Second, DNN models are not identifiable, because the empirical risk minimization (ERM) is non-convex with high dimensionality. Given the ERM being non-convex, the DNN training is highly sensitive to the initialization \cite{He2015,Glorot2010}. With different initializations, the DNN model can end with local minima or saddle points, rather than the global optimum \cite{Goodfellow2016,Dauphin2014}. This issue does not happen in the classical MNL models, because the ERM of the MNL models is globally convex \cite{Boyd2004}. Decades ago, model non-identification was one reason why DNNs were discarded \cite{LeCun2015}. Nowadays, however, researchers argue that some high quality local minima are acceptable, and the global minimum in the training may be irrelevant since the global minimum tends to overfit \cite{Choromanska2015}. This problem of model non-identification indicates that each training of DNNs can lead to very different models, even conditioned on the fixed hyperparameters and training samples. Importantly these trained DNNs may have very similar prediction performance, creating difficulties for researchers to choose the final model for interpretation. 

Third, the choice probability functions in DNNs can be locally irregular because their gradients can be exploding or the functions themselves are non-monotonic, both of which are discussed under the robust DNN framework. When the gradients of choice probability functions are exploding, it is very simple to find an adversarial input $x'$, which is $\epsilon$-close to the initial $x$ ($ ||x' - x||_p \leq \epsilon$) but is wrongly predicted to be a label different from the initial $x$ with high confidence. This type of system is not robust because they can be easily fooled by the adversarial example $x'$. In fact, it has been found that DNNs lack robustness \cite{Nguyen2015,Szegedy2014}. With even a small $\epsilon$ perturbation introduced to an input image $x$, DNNs label newly generated image $x'$ to the wrong category with extremely high confidence, when the correct label should be the same as the initial input image $x$ \cite{Szegedy2014,Goodfellow2015}. Therefore, the lack of robustness in DNNs implies the locally irregular patterns of the choice probability functions and the gradients, which are the key information for DNN interpretation. 

\section{Model}
\label{s:3}
\subsection{DNNs for Choice Analysis}
\begin{figure}
\centering
\scalebox{.7}{



\def\layersep{1.5cm}
\begin{tikzpicture}[shorten >=1pt, ->, draw=black!50, node distance=\layersep]
    \tikzstyle{every pin edge}=[<-, shorten <=1pt]
    \tikzstyle{neuron}=[circle,fill=black!25,minimum size=6pt,inner sep=0pt]
    \tikzstyle{input neuron}=[neuron, fill=black!50];
    \tikzstyle{output neuron}=[neuron, fill=red!50];
    \tikzstyle{hidden neuron}=[neuron, fill=blue!50];
    \tikzstyle{annot} = [text width=3cm, text centered]

    \foreach \name / \y in {1,...,4}
        \node[input neuron, pin=left:$x_{\y}$] (I-\name) at (0 cm,-\y cm) {};

    \filldraw [black!50] (0, -4.5cm) circle (2pt);
    \filldraw [black!50] (0, -5.0cm) circle (2pt);
    \filldraw [black!50] (0, -5.5cm) circle (2pt);
    \node[input neuron, pin=left:$x_{20}$] (I-5) at (0 cm,-6cm) {};

    \foreach \name / \y in {1,...,5}
        \path[yshift=0.5cm]
            node[hidden neuron] (H1-\name) at (\layersep, -\y cm) {};

    \filldraw [blue!50] (\layersep, -5.0cm) circle (2pt);
    \filldraw [blue!50] (\layersep, -5.5cm) circle (2pt);
    \filldraw [blue!50] (\layersep, -6.0cm) circle (2pt);
    \node[hidden neuron] (H1-6) at (\layersep,-6.5cm) {};

    \foreach \name / \y in {1,...,5}
        \path[yshift=0.5cm]
            node[hidden neuron] (H2-\name) at (2*\layersep, -\y cm) {};
    \filldraw [blue!50] (2*\layersep, -5.0cm) circle (2pt);
    \filldraw [blue!50] (2*\layersep, -5.5cm) circle (2pt);
    \filldraw [blue!50] (2*\layersep, -6.0cm) circle (2pt);
    \node[hidden neuron] (H2-6) at (2*\layersep,-6.5cm) {};

    \foreach \name / \y in {1,...,5}
        \path[yshift=0.5cm]
            node[hidden neuron] (H3-\name) at (3*\layersep, -\y cm) {};
    \filldraw [blue!50] (3*\layersep, -5.0cm) circle (2pt);
    \filldraw [blue!50] (3*\layersep, -5.5cm) circle (2pt);
    \filldraw [blue!50] (3*\layersep, -6.0cm) circle (2pt);
    \node[hidden neuron] (H3-6) at (3*\layersep,-6.5cm) {};

    \foreach \name / \y in {1,...,5}
        \path[yshift=0.5cm]
            node[hidden neuron] (H4-\name) at (4*\layersep, -\y cm) {};
    \filldraw [blue!50] (4*\layersep, -5.0cm) circle (2pt);
    \filldraw [blue!50] (4*\layersep, -5.5cm) circle (2pt);
    \filldraw [blue!50] (4*\layersep, -6.0cm) circle (2pt);
    \node[hidden neuron] (H4-6) at (4*\layersep,-6.5cm) {};

    \foreach \name / \y in {1,...,5}
        \path[yshift=0.5cm]
            node[hidden neuron] (H5-\name) at (5*\layersep, -\y cm) {};
    \filldraw [blue!50] (5*\layersep, -5.0cm) circle (2pt);
    \filldraw [blue!50] (5*\layersep, -5.5cm) circle (2pt);
    \filldraw [blue!50] (5*\layersep, -6.0cm) circle (2pt);
    \node[hidden neuron] (H5-6) at (5*\layersep,-6.5cm) {};

    \foreach \name / \y in {1,...,5}
        \path[yshift=0.5cm]
            node[hidden neuron] (H6-\name) at (6*\layersep, -\y cm) {};
    \filldraw [blue!50] (6*\layersep, -5.0cm) circle (2pt);
    \filldraw [blue!50] (6*\layersep, -5.5cm) circle (2pt);
    \filldraw [blue!50] (6*\layersep, -6.0cm) circle (2pt);
    \node[hidden neuron] (H6-6) at (6*\layersep,-6.5cm) {};

    \foreach \name / \y in {1,...,5}
        \path[yshift=0.5cm]
            node[hidden neuron] (H7-\name) at (7*\layersep, -\y cm) {};
    \filldraw [blue!50] (7*\layersep, -5.0cm) circle (2pt);
    \filldraw [blue!50] (7*\layersep, -5.5cm) circle (2pt);
    \filldraw [blue!50] (7*\layersep, -6.0cm) circle (2pt);
    \node[hidden neuron] (H7-6) at (7*\layersep,-6.5cm) {};

    \foreach \name / \y in {1,...,5}
        \node[output neuron, pin={[pin edge={->}]right:$s_{\y}$}] (O-\name) at (8*\layersep, -\y cm - 0.5cm) {};


    \foreach \source in {1,...,5}
        \foreach \dest in {1,...,6}
            \path (I-\source) edge (H1-\dest);

    \foreach \source in {1,...,6}
        \foreach \dest in {1,...,6}
            \path (H1-\source) edge (H2-\dest);

    \foreach \source in {1,...,6}
        \foreach \dest in {1,...,6}
            \path (H2-\source) edge (H3-\dest);

    \foreach \source in {1,...,6}
        \foreach \dest in {1,...,6}
            \path (H3-\source) edge (H4-\dest);

    \foreach \source in {1,...,6}
        \foreach \dest in {1,...,6}
            \path (H4-\source) edge (H5-\dest);

    \foreach \source in {1,...,6}
        \foreach \dest in {1,...,6}
            \path (H5-\source) edge (H6-\dest);

    \foreach \source in {1,...,6}
        \foreach \dest in {1,...,6}
            \path (H6-\source) edge (H7-\dest);

    \foreach \source in {1,...,6}
        \foreach \dest in {1,...,5}
            \path (H7-\source) edge (O-\dest);


\end{tikzpicture}
}
\caption{A feedforward DNN architecture (7 hidden layers * 100 neurons)}
\label{fig:nn_arch}
\end{figure}
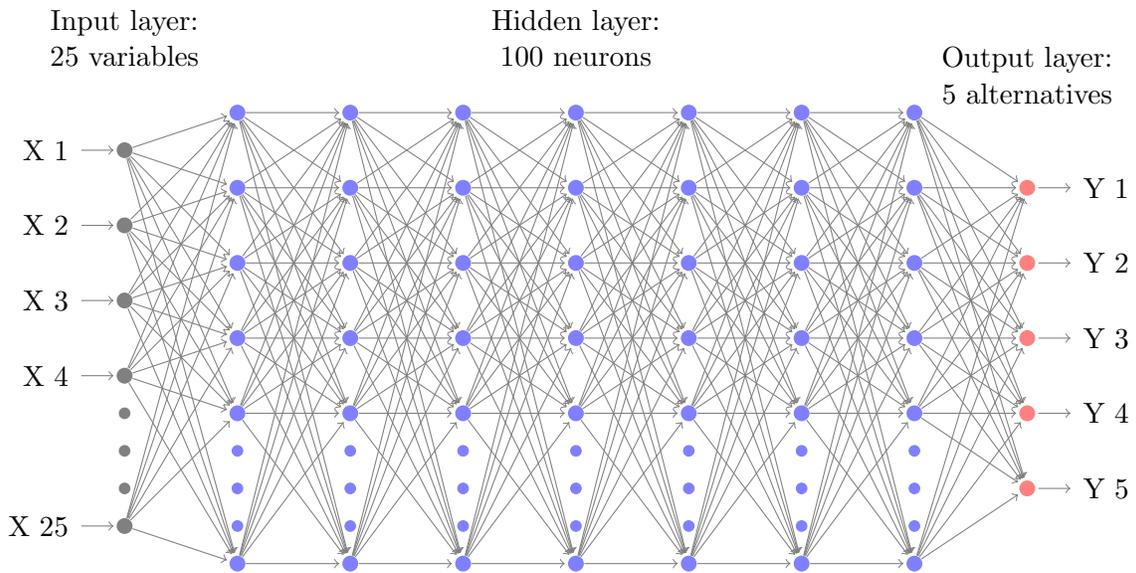

\noindent
DNNs can be applied to choice analysis. Let $s^*_k(x_i)$ denote the true probability of individual $i$ choosing alternative $k$ out of $[1, 2, ..., K]$ alternatives, with $x_i$ denoting the input variables: $s^*_k(x_i): {R}^d \rightarrow [0, 1]$. Individual $i$'s choice $y_{i} \in \{0,1 \}^K$ is sampled from a multinomial random variable with $s^*_k(x_i)$ probability of choosing $k$. With DNNs applied to choice analysis, the choice probability function is:
\begin{flalign} \label{eq:choice_prob_dnn}
s_k(x_i) = \frac{e^{V_{ik}}}{ \sum_j e^{V_{ij}}}
\end{flalign}

\noindent
in which $V_{ij}$ and $V_{ik}$ are the $j$th and $k$th inputs into the Softmax activation function of DNNs. $V_{ik}$ takes the layer-by-layer form:
\begin{flalign} \label{eq:util_dnn}
V_{ik} = (g_m^k \circ g_{m-1} ... \circ g_2 \circ g_1)(x_i)
\end{flalign}

\noindent
where each $g_l(x) = ReLU(W_l x + b_l)$ is the composition of linear and rectified linear unit (ReLU) transformation; $g_m^k$ represents the transformation of the last hidden layer into the utility of alternative $k$; and $m$ is the total number of layers in a DNN. Figure \ref{fig:nn_arch} visualizes a feedforward DNN architecture with $20$ input variables, $5$ output alternatives, and $7$ hidden layers. The grey nodes represent the input variables; the blue ones represent the hidden layers; and the red ones represent the Softmax activation function. The layer-by-layer architecture in Figure \ref{fig:nn_arch} reflects the compositional structure of Equation \ref{eq:util_dnn}.

The inputs into the Softmax layers in DNNs can be treated as utilities, the same as those in the classical DCMs. This utility interpretation in DNNs is shown by the Lemma 2 in McFadden (1974) \cite{McFadden1974}, which implies that the Softmax activation function is equivalent to a random utility term with Gumbel distribution under the random utility maximization (RUM) framework. Hence DNNs and MNL models are both under the RUM framework, and their difference only resides in the utility specifications. In other words, the inputs into the last Softmax activation function of DNNs can be interpreted as utilities; the outputs from the Softmax activation function are choice probabilities; the transformation before this Softmax function can be seen as the specification of utility functions; and the Softmax activation function can be seen as the comparison of utility values. 

Despite their similarity, DNNs are a much more generic model family than MNL models, and this relationship can be understood from various perspectives. The universal approximator theorem developed in the 1990s indicates that a neural network with only one hidden layer is asymptotically a universal approximator when the width becomes infinite \cite{Cybenko1989,Hornik1989,Hornik1991}. Recently this asymptotic perspective leads to a more non-asymptotic question, asking why depth is necessary when a wide and shallow neural network is already powerful enough. It can be shown that DNNs can approximate functions with an exponentially smaller number of neurons than a shallow neural network in many settings \cite{Cohen2016,Rolnick2017,Poggio2017}. In other words, DNNs can be treated as an efficient universal approximator, thus being much more generic than the MNL model, which is a shallow neural network with zero hidden layers. However, a more generic model family leads to both smaller approximation errors and large estimation errors. Since the out-of-sample prediction error equals to the sum of the approximation and estimation errors, DNNs do not necessarily outperform MNL models from a theoretical perspective. The major challenge of DNNs is its large estimation error, which is associated with its extraordinary approximation power. To find the best balance between the approximation and estimation errors, the procedure of hyperparameter searching needs to be used since the hyperparameters, such as the DNNs' depth and width, control the model complexity. A brief theoretical proof about the large estimation error of DNNs is available in Appendix I. More detailed discussions are available in the recent studies from statistical learning theory \cite{Vershynin2018,Wainwright2019,Golowich2017,Neyshabur2015,Bartlett2002,Ledoux2013,Bartlett2006}. 

\subsection{Computing Economic Information From DNNs}
\noindent
The utility interpretation in DNNs enables us to derive all the economic information traditionally obtained from DCMs. With $\hat{V}_k(x_i)$ denoting the estimated utility of alternative $k$ and $\hat{s}_k(x_i)$ the estimated choice probability function, Table \ref{table:dnn_econ_info} summarizes the formula of computing the economic information, which is sorted into two categories. Choice probabilities, choice predictions, market share, substitution patterns, and social welfare are derived by using functions (either choice probability or utility functions). Probability derivatives, elasticities, MRS, and VOTs are derived from the gradients of choice probability functions. This differentiation is owing to the the different theoretical properties between functions and their gradients \footnote{The uniform convergence proof is possible for the estimated functions, while it is much harder for the gradients because the estimated functions may not be even differentiable.}. The formula in Table \ref{table:dnn_econ_info} can be applied to both DNNs and MNLs, but the MNLs have a further explicit parametric form for each piece of economic information while DNNs don't \cite{Train2009}.

\begin{table}[ht]
\centering
\caption{Formula to compute economic information in both DNNs and DCMs; F stands for function, GF stands for the gradients of functions.}
\resizebox{1.0\linewidth}{!}{ 
\begin{tabular}{p{0.4\linewidth} | P{0.4\linewidth} | P{0.15\linewidth}}
\toprule
\hline
\textbf{Economic Information} & \textbf{Formula in DNNs} & \textbf{Categories} \\
\hline
Choice probability & $\hat{s}_k(x_i)$ & F \\
\hline
Choice prediction & $\underset{k}{\argmax} \ \hat{s}_k(x_i)$ & F \\
\hline
Market share & $\sum_i \hat{s}_k(x_i)$ & F \\
\hline
Substitution pattern between alternatives $k_1$ and $k_2$ & $\hat{s}_{k_1}(x_i) / \hat{s}_{k_2}(x_i)$ & F \\
\hline
Social welfare & $\sum_i \frac{1}{\alpha_i} \log (\sum_{k=1}^K e^{\hat{V}_{ik}}) + C$ & F \\
\hline
Change of social welfare & $\sum_i \frac{1}{\alpha_i} \big[ \log (\sum_{k=1}^K e^{\hat{V}^1_{ik}}) - \log (\sum_{k=1}^K e^{\hat{V}^0_{ik}}) \big]$ & F \\
\hline
Probability derivative of alternative $k$ w.r.t. $x_{ij}$ & $\partial \hat{s}_k(x_i)/\partial x_{ij}$ & GF \\
\hline
Elasticity of alternative $k$ w.r.t. $x_{ij}$ & $\partial \hat{s}_k(x_i)/\partial x_{ij} \times x_{ij}/\hat{s}_k(x_i) $ & GF \\
\hline
Marginal rate of substitution between $x_{ij_1}$ and $x_{ij_2}$ & $ - \frac{\partial \hat{s}_k(x_i)/\partial x_{ij_1}}{\partial \hat{s}_k(x_i)/\partial x_{ij_2}}$ & GF \\
\hline
VOT ($x_{ij_1}$ is time and $x_{ij_2}$ is monetary value) & $ - \frac{\partial \hat{s}_k(x_i)/\partial x_{ij_1}}{\partial \hat{s}_k(x_i)/\partial x_{ij_2}}$ & GF \\
\hline
\bottomrule
\end{tabular}
} 
\label{table:dnn_econ_info}
\end{table}

This process of interpreting economic information from DNNs is significantly different from the classical DCMs for the following reasons. In DNNs, the economic information is directly computed by using the \textit{full functions} $\hat{s}_k(x_i)$ and $\hat{V}_k(x_i)$, rather than \textit{individual parameters} $\hat{w}$. This focus on functions rather than individual parameters in DNNs is inevitable because the non-convex and high-dimensional DNN training leads to unstable parameter estimates, while MNL has the same estimate in every training owing to the convexity of its empirical risk minimization. This focus on the full functions is also consistent with other studies concerning the interpretation of DNNs: a large number of recent studies focused on the full functions of DNNs for interpretation, while none focused on individual neurons/parameters \cite{Montavon2018,Hinton2015,Baehrens2010,Ross2018}. Hence the DNN interpretation can be seen as an end-to-end mechanism without involving the individual parameters as an intermediate process. In addition, the interpretation of DNNs is a prediction-driven process: the economic information is generated in a post-hoc manner after a model is trained to be highly predictive. This prediction-driven interpretation takes advantage of DNNs' capacity of automatic feature learning, and it is also in contrast to the classical DCMs that rely on handcrafted utility functions. This prediction-driven interpretation is based on the belief that ``when predictive quality is (consistently) high, some structure must have been found'' \cite{Mullainathan2017}.

\subsection{MNLs for Choice Analysis}
\noindent
The classical MNL with linear specification is used as the reference point to the DNNs in our empirical experiments. The utility function in the MNL models is shown as the following:
\begin{flalign} \label{eq:util_mnl}
& V_{ik} = \beta_{0,k} + \beta_{x,k}^T x_{ik} + \beta_{z,k}^T z_i, \ \ as \ \ k \neq ref \\
& V_{ik} = \beta_{x,k}^T x_{ik}, \ \ as \ \ k = ref 
\end{flalign}
\noindent
in which $V_{ik}$ is the deterministic utility value for alternative $k$; $\beta_{0,k}$ represents the alternative-specific constant for alternative $k$; $\beta_{x,k}$ represents the parameters for the alternative-specific variables $x_{ik}$; $\beta_{z,k}$ represents the parameters for the individual-specific variables $z_i$; $ref$ represents the reference alternative. For parameter identification, the utility specification is different depending on whether the alternative is used as the reference. This formulation is the simplest specification that guarantees the parameter identification in choice modeling. A more generic form is $V_{ik} = \beta_{0,k} + \beta_{x,k}^T \phi_x(x_{ik}) + \beta_{z,k}^T \phi_z(z_i)$, in which $\phi_x$ and $\phi_z$ represent the functions for feature transformation, such as quadratic and log transformation. This study uses the linear specification for two reasons. First for fairness, both MNL and DNNs use the linear inputs, so their comparison is not biased. Second for simplicity, while we use only linear specification, future studies can compare DNNs to the MNLs with feature transformations.

\section{Setup of Experiments}
\noindent
The experiments use two groups of DNN models, referred to as Random-DNNs and Opt-DNNs. Random-DNNs are those DNNs trained with varying hyperparameters \textit{randomly} chosen within a prespecified hyperparameter space, and Opt-DNNs refer to the repeated trainings of the DNNs with the \textit{fixed} hyperparameters that perform the best in the Random-DNNs. 

\subsection{Random-DNNs: Hyperparameter Training}
\label{s:random_training}
\noindent
The group of Random-DNNs is constructed by randomly exploring a pre-specified hyperparameter space and using the sampled set of hyperparameters for each DNN training \cite{Bergstra2012}. The hyperparameter space consists of the architectural hyperparameters, including the depth and width of DNNs; and the regularization hyperparameters, including $L_1$ and $L_2$ penalty constants, and dropout rates. 100 sets of hyperparameters are randomly generated for comparison. The details of the hyperparameter space is available in Appendix II. Besides the hyperparameters varying across the $100$ models, all the DNN models share certain fixed components, including ReLU activation functions in the hidden layers, Softmax activation function in the last layer, Gloret initialization, and Adam optimization, following the standard practice \cite{Goodfellow2016,Geron2017}. Formally, the hyperparameter searching is formulated as
\begin{equation}
\setlength{\jot}{2pt} \label{eq:hyper_searching_2}
  \begin{aligned}
  \hat{w}_h = \underset{w_h \in \{w_h^{(1)},w_h^{(2)}, ..., w_h^{(S)} \} }{\argmin} \ \underset{w}{\argmin} \ L(w, w_h)
  \end{aligned}
\end{equation}

\noindent
where $L(w, w_h)$ is the empirical risk function that the DNN training aims to minimize, $w$ represents the parameters in a DNN architecture, $w_h$ represents the hyperparameter, $w_h^{(s)}$ represents one set of hyperparameters randomly sampled from the hyperparameter space, and $\hat{w}_h$ is the chosen hyperparameter with the highest prediction accuracy. Besides this baseline random searching, other approaches can be used for hyperparameter training, such as reinforcement learning or Bayesian methods \cite{Snoek2015,Zoph2016}, which are beyond the scope of our study.

\subsection{Opt-DNNs: Training with Fixed Hyperparameters}
\label{s:fixed_training}
\noindent
After the hyperparameter searching, we examine a set of optimum hyperparameters that lead to the highest prediction accuracy. By using the same training set and the fixed set of optimum hyperparameters, we train the DNN models another $100$ times to construct the group of Opt-DNNs. Each training seeks to minimize the empirical risk conditioned on the fixed hyperparameters, formulated as following.
\begin{equation}
\setlength{\jot}{2pt} \label{eq:erm}
  \begin{aligned}
  \underset{w}{\min} \ L(w, \hat{w}_h) = \underset{w}{\min} \ \frac{1}{N} \sum_{i=1}^{N} l(y_i, s_{k}(x_i; w, \hat{w}_h)) + \gamma ||w||_p
  \end{aligned}
\end{equation}

\noindent
where $w$ represents the parameters; $\hat{w}_h$ represents the best hyperparameters; $l()$ is the loss function, typically the cross-entropy loss; $N$ is the sample size. $\gamma ||w||_p$ represents $L_p$ penalty ($||w||_p = (\sum_j (w_j)^p)^{\frac{1}{p}} $), and $L_1$ (LASSO) and $L_2$ (Ridge) penalties are the two specific cases of $L_p$ penalties. Note that DNNs have the model non-identification challenge because the objective function in Equation \ref{eq:erm} is not globally convex. DNNs have the local irregularity challenge because this optimization over the \textit{global} prediction risks is insufficient to guarantee the \textit{local} fidelity. The two issues will be demonstrated in more details in Section \ref{s:4}.

\subsection{Datasets}
\noindent
Our experiments rely on two datasets: a stated preference survey conducted in Singapore and a revealed preference data in London. The Singapore data set was collected by the authors, with the help of a professional survey company Qualtrics in July 2017. The survey started with asking the respondents to report their postal codes of their home and working locations and their current travel mode. From respondents' home (origin) and work (destination) locations, we computed the walking time, waiting time, in-vehicle travel time, and travel cost of each travel mode for each individual’s commute trip using Google Maps API. The information was then used to automatically generate the stated preference section, which was the bulk of the questionnaire. The respondents were asked to choose among five travel modes: walking, public transit, driving, ride sharing, and shared autonomous vehicles, with varying values of price and travel time. In the end, respondents reported socioeconomic information such as gender, education, and income. The London data set was publicly provided in Hillel, et al. (2018) \cite{Hillel2018}, in which the authors constructed a new data set based on the London Travel Demand Survey (LTDS) by combining the individual trip records and the trip trajectories along the mode alternatives. The authors started with LTDS, removed the trips that had the same origin-destination post codes, assigned each trip to one of four main travel modes (walking, cycling, public transport, and driving), and simplified the trip purposes to five purposes (B, HBW, HBE, HBO, and NHBO). The authors then augmented travel time and price information to the initial LTDS by using Google Map API and Oyster cards. 

To understand the impacts of different sample sizes and contexts, we constructed three datasets from these two initial data sets for our experiments: (1) the SGP dataset with the full $8,418$ observations (8K-SGP), (2) the LD dataset with the full $81,086$ observations (80K-LD), and (3) the LD dataset with randomly sampled $8,000$ observations (8K-LD). The comparison between the 8K-SGP and 8K-LD datasets reveals the effect of two contexts, and the comparison between the 80K-LD and 8K-LD datasets reveals the effect of the sample size. All the datasets are split into training and testing sets by using the default random seed in the Numpy module in Python. The sample sizes of training and testing sets in 8K-SGP are $7,015$ and $1,403$, and that in 80K-LD are $72,977$ and $8,109$. In both datasets, the choice variable $y$ is travel mode choice, including five alternatives (walking, taking public transit, ride sharing, using an autonomous vehicle, and driving) in the SGP dataset and four alternatives (walking, cycling, driving, and using public transit) in the LD dataset. The explanatory variables include $20$ individual-specific and alternative-specific variables in the SGP dataset and $14$ variables in the LD dataset. Please refer to Appendix III for the summary statistics of the two data sets.

\section{Experimental Results}
\label{s:4}
\noindent
This section shows that it is feasible to extract all the economic information from DNNs without using individual parameters, and that by using large sample, hyperparameter searching, model ensemble, and regularization methods, it is possible to extract reliable economic information. We will first present prediction accuracy, then the function-based interpretation for choice probabilities, substitution patterns of alternatives, market share, and social welfare, and lastly the gradient-based interpretation for probability derivatives, elasticities, VOT, and heterogeneous preferences. This section summarizes two groups of DNN models (Opt-DNNs and Random-DNNs) and the linear MNL model, applied to the 8K-SGP, 80K-LD, and 8K-LD datasets.

\subsection{Prediction Accuracy}
\begin{figure}[ht]
\captionsetup[subfigure]{justification=centering}
\centering
\subfloat[Opt-DNNs (8K-SGP)]{\includegraphics[height=0.23\linewidth]{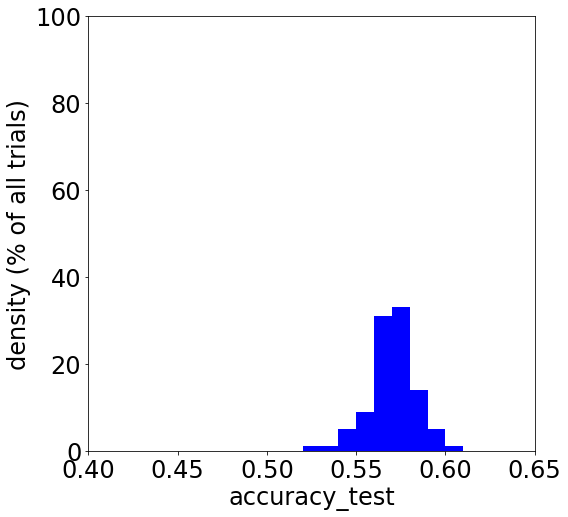}\label{sfig:pred_accuracy_opt_sgp}}
\subfloat[Random-DNNs (8K-SGP)]{\includegraphics[height=0.23\linewidth]{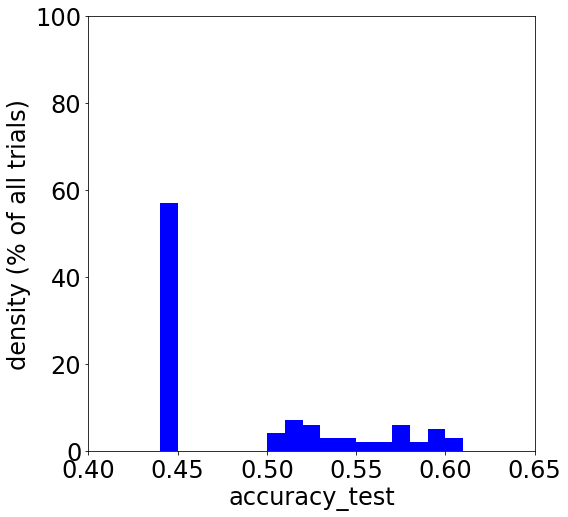}\label{sfig:pred_accuracy_random_sgp}}
\subfloat[MNL (8K-SGP)]{\includegraphics[width=0.25\linewidth]{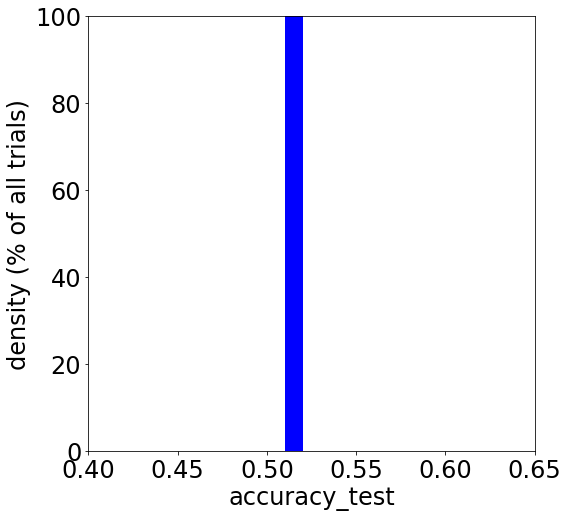}\label{sfig:pred_accuracy_mnl_sgp}} \\
\subfloat[Opt-DNNs (8K-LD)]{\includegraphics[height=0.23\linewidth]{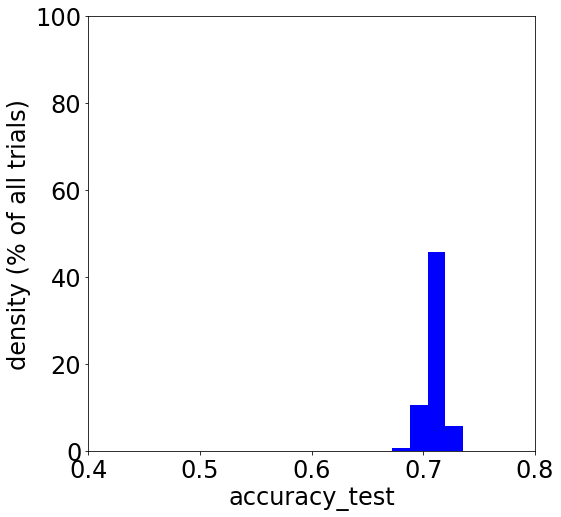}\label{sfig:pred_accuracy_opt_8k_ld}}
\subfloat[Opt-DNNs (80K-LD)]{\includegraphics[height=0.23\linewidth]{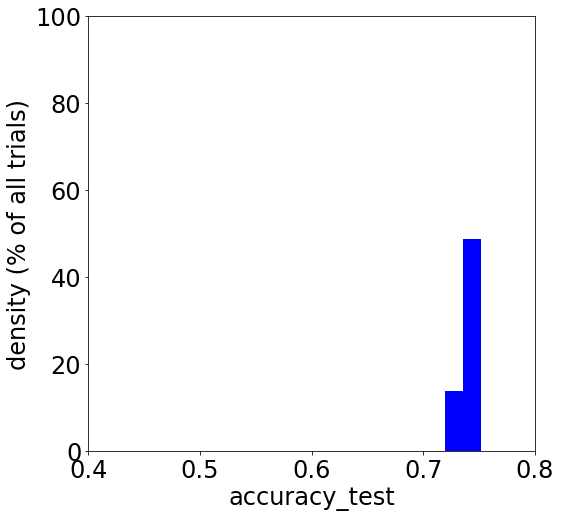}\label{sfig:pred_accuracy_opt_80k_ld}}
\subfloat[Random-DNNs (80K-LD)]{\includegraphics[height=0.23\linewidth]{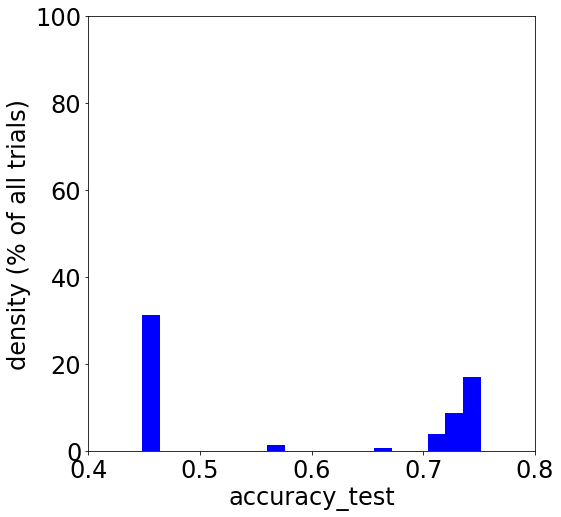}\label{sfig:pred_accuracy_random_80k_ld}}
\subfloat[MNL (80K-LD)]{\includegraphics[width=0.25\linewidth]{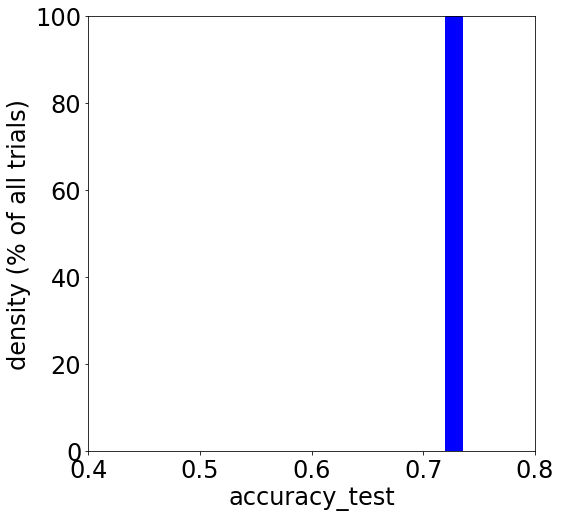}\label{sfig:pred_accuracy_mnl_80k_ld}} \\
\caption{Histograms of the prediction accuracy in seven scenarios (100 trainings for each model group)} 
\label{fig:pred_accuracy}
\end{figure}

\noindent
Figures \ref{sfig:pred_accuracy_opt_sgp}-\ref{sfig:pred_accuracy_mnl_80k_ld} reveal three findings. First, Opt-DNNs on average outperform the MNL models by about 2 to 8 percentage points prediction accuracy, which is consistent with the previous studies that found the outperformance of DNNs over MNL \cite{Rao1998,Nijkamp1996,Karlaftis2011}. Second, choosing the correct hyperparameter plays a critical role in improving the model performance of DNNs, as shown by the higher prediction accuracy of the Opt-DNNs than the Random-DNNs in both 8K-SGP and 80K-LD datasets. Third, larger sample size improves the prediction accuracy of DNNs, as shown by comparing Figures \ref{sfig:pred_accuracy_opt_8k_ld} and \ref{sfig:pred_accuracy_opt_80k_ld}. 

\subsection{Function-Based Interpretation}
\subsubsection{Choice Probability Functions}
\label{sec:cpf}
\begin{figure}[ht]
\captionsetup[subfigure]{justification=centering}
\centering
\subfloat[Opt-DNNs (8K-SGP)]{\includegraphics[width=0.23\linewidth]{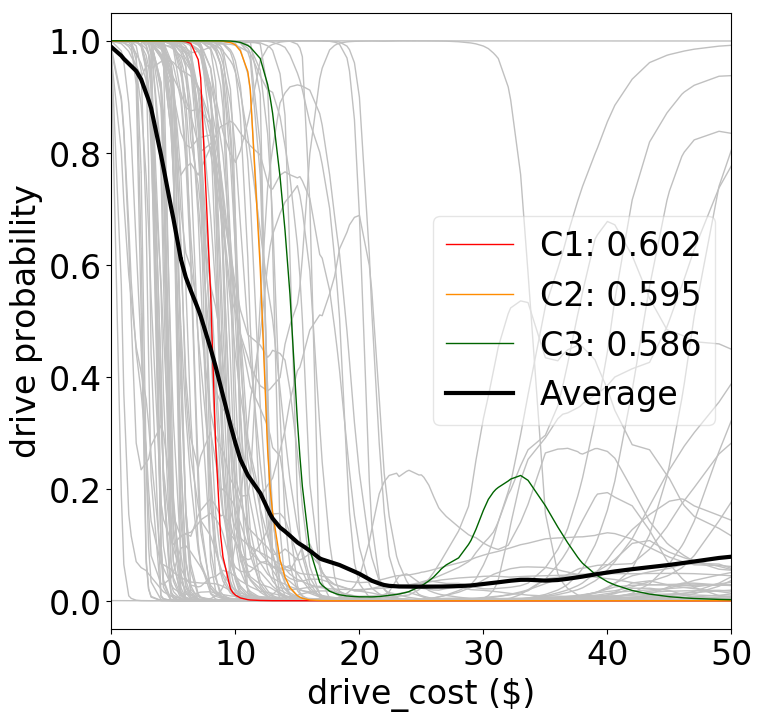}\label{sfig:prob_opt_8k_sgp}} 
\subfloat[Random-DNNs (8K-SGP)]{\includegraphics[width=0.23\linewidth]{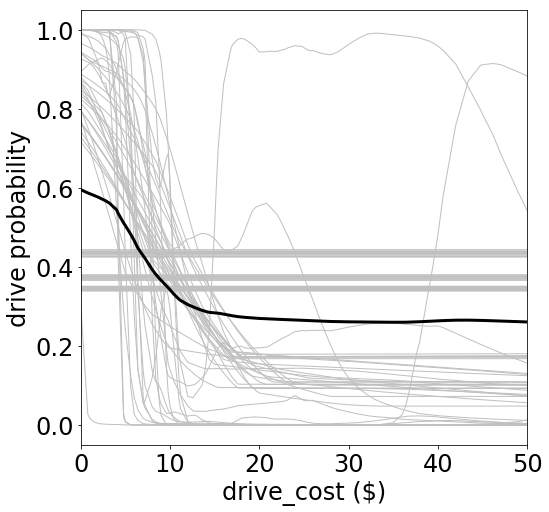}\label{sfig:prob_random_8k_sgp}}
\subfloat[MNL (8K-SGP)]{\includegraphics[width=0.23\linewidth]{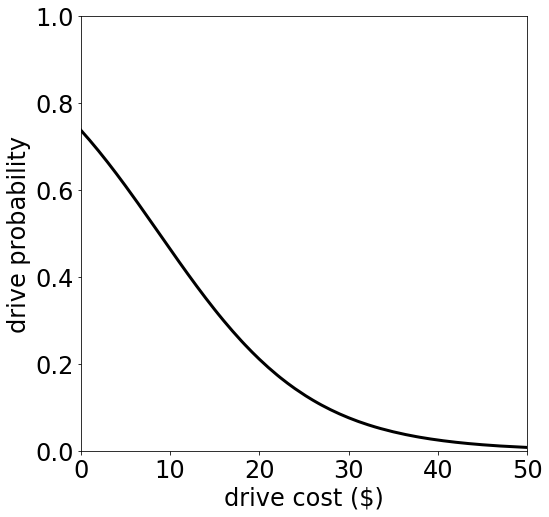}\label{sfig:prob_mnl_8k_sgp}}\\
\subfloat[Opt-DNNs (8K-LD)]{\includegraphics[width=0.23\linewidth]{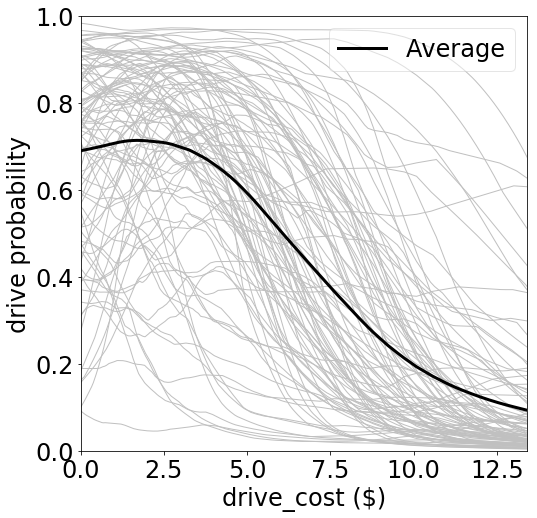}\label{sfig:prob_opt_8k_ld}}
\subfloat[Opt-DNNs (80K-LD)]{\includegraphics[width=0.23\linewidth]{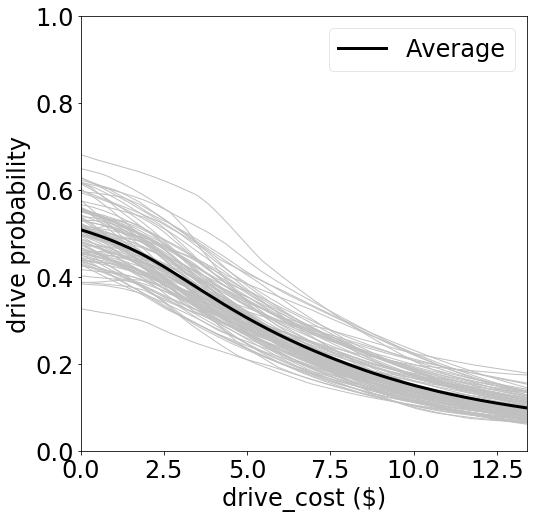}\label{sfig:prob_opt_80k_ld}} 
\subfloat[Random-DNNs (80K-LD)]{\includegraphics[width=0.23\linewidth]{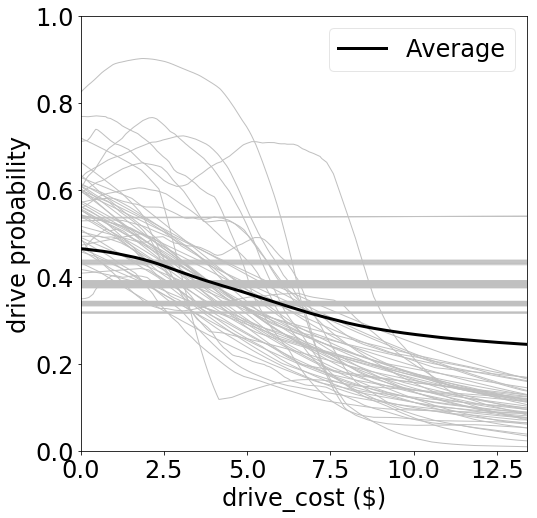}\label{sfig:prob_random_80k_ld}}
\subfloat[MNL (80K-LD)]{\includegraphics[width=0.23\linewidth]{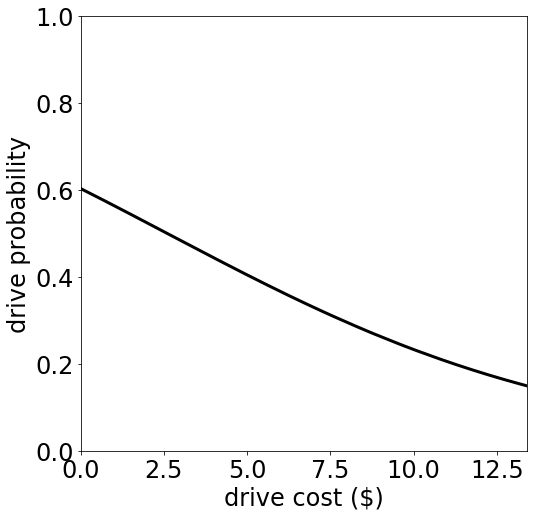}\label{sfig:prob_mnl_80k_ld}}\\
\caption{Driving probability functions with driving costs (100 trainings for each model group)}
\label{fig:prob}
\end{figure}

\noindent
The choice probability functions are visualized in Figure \ref{fig:prob}. Since the inputs of the choice probability functions $s(x)$ have high dimensions, the $s(x)$ is visualized by computing the driving probability with varying only the driving cost, holding all the other variables constant at the sample mean. Each light grey curve in Figures \ref{sfig:prob_opt_8k_sgp}-\ref{sfig:prob_mnl_80k_ld} represents one individual training result, and the dark curve is the ensemble of all $100$ models. In Figures \ref{sfig:prob_mnl_8k_sgp} and \ref{sfig:prob_mnl_80k_ld}, only one training result is visualized because the MNL training has no variation.

Figure \ref{fig:prob} demonstrates the power of DNNs being able to automatically learn the choice probability functions in both large and small sample sizes. With large sample size, the choice probability functions in Figure \ref{sfig:prob_opt_80k_ld} are highly concentrated and are quite reasonable, similar to the pattern in Figure \ref{sfig:prob_mnl_80k_ld}. Even with small sample size, the majority of the choice probability functions in Figures \ref{sfig:prob_opt_8k_sgp} and \ref{sfig:prob_opt_8k_ld} are roughly decreasing. The choice probabilities are high when the driving cost is close to zero, while they become much smaller when the driving cost increases. This roughly decreasing pattern is reasonable from a behavioral perspective. In comparison to the choice probability functions of MNL (Figure \ref{sfig:prob_mnl_8k_sgp}), the choice probability functions of the Opt-DNNs in Figure \ref{sfig:prob_opt_8k_sgp} are richer and more flexible. However, the caveat is that the DNN choice probability functions may be too flexible to reflect the true behavioral mechanism in the small sample, owing to three theoretical challenges.

First, the large variation of the Random-DNNs in Figures \ref{sfig:prob_random_8k_sgp} and \ref{sfig:prob_random_80k_ld} reveal that DNN models are sensitive to the choice of hyperparameters. With different hyperparameters, some of Random-DNNs' choice probability functions are simply flat without revealing any useful information, while others are similar to Opt-DNNs with reasonable patterns. This challenge can be mitigated by hyperparameter searching. For example, the Opt-DNNs can reveal more reasonable economic information than the Random-DNNs because the Opt-DNNs use specific architectural and regularization hyperparameters, chosen from the results of hyperparameter searching based on their high prediction accuracy. 

Second, the large variation of the individual Opt-DNN trainings (Figures \ref{sfig:prob_opt_8k_sgp}, \ref{sfig:prob_opt_8k_ld}, and \ref{sfig:prob_opt_80k_ld}) reveal the challenge of model non-identification. Given that the 100 trainings are conditioned on the same training data and the same set of hyperparameters, the variation across the Opt-DNNs can only be attributed to the model non-identification issue, or more specifically, the optimization difficulties in minimizing the non-convex risk function of DNNs. As DNNs' risk function is non-convex, different model trainings can converge to very different local minima or saddle points. Whereas these local minima have similar prediction accuracy, it brings difficulties to the model interpretation since the functions learnt from different local minima are different. For example in Figure \ref{sfig:prob_opt_8k_sgp}, the three individual training results (C1, C2, and C3) have very similar out-of-sample prediction accuracy ($60.2 \%$, $59.5 \%$, and $58.6 \%$); however, their corresponding choice probability functions are very different. In fact, the majority of the $100$ individual trainings have quite similarly high prediction accuracy, whereas their choice probability functions differ from each other. On the other side, the choice probability function averaged over the 100 trainings of the Opt-DNNs is more stable than individual ones, demonstrating the importance of model ensemble in controlling generalization errors and smoothing choice probability functions. 

Third, the shapes of the individual choice probability curves show the local irregularity of the choice probability functions. Let's take Figure \ref{sfig:prob_opt_8k_sgp} as an example. Some choice probability functions can be sensitive to the small change of input values: the probability of choosing driving in C1 drops from $96.6\%$ to $7.8\%$ as the driving cost increases from $\$7$ to $\$9$, indicating a locally exploding gradient. This phenomenon of exploding gradients is acknowledged in the robust DNN discussions, because exploding gradients render a system vulnerable \cite{Ross2017,Ross2018}. Many training results present a non-monotonic pattern. For example, C3 represents a counter-intuitive case where the probability of driving starts to increase dramatically as the driving costs are larger than $\$25$. However, it is worth noting that this local irregularity problem can be mitigated with model ensemble and larger sample size: the choice probability functions in Figure \ref{sfig:prob_opt_80k_ld} are much more regular than those in Figure \ref{sfig:prob_opt_8k_ld}.

\subsubsection{Substitution Pattern of Alternatives}
\label{sec:spa}
\begin{figure}[ht]
\captionsetup[subfigure]{justification=centering}
\centering
\subfloat[Opt-DNNs (8K-SGP)]{\includegraphics[width=0.23\linewidth]{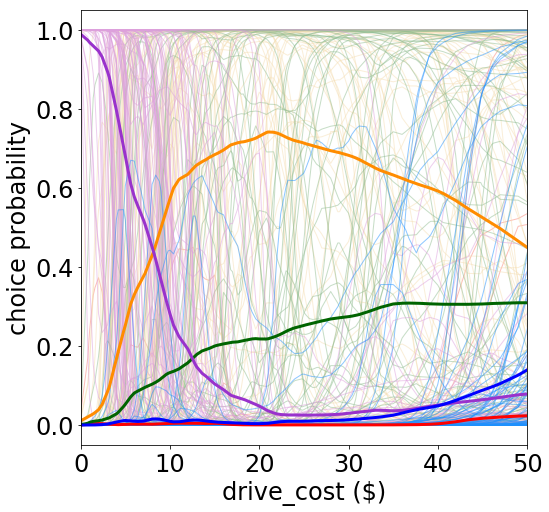}\label{sfig:substitution_opt_8k_sgp}}
\subfloat[Random-DNNs (8K-SGP)]{\includegraphics[width=0.23\linewidth]{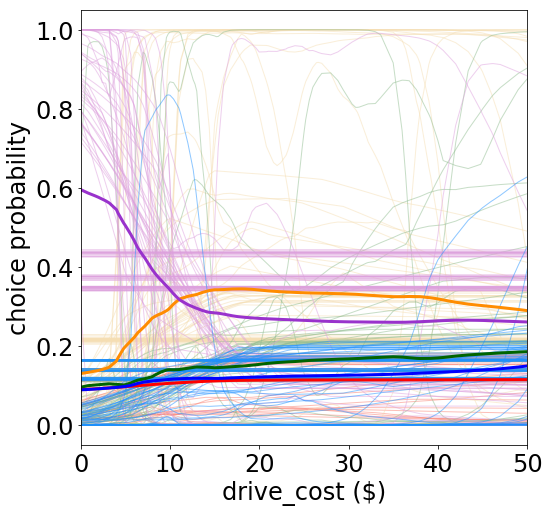}\label{sfig:substitution_random_8k_sgp}}
\subfloat[MNL (8K-SGP)]{\includegraphics[width=0.23\linewidth]{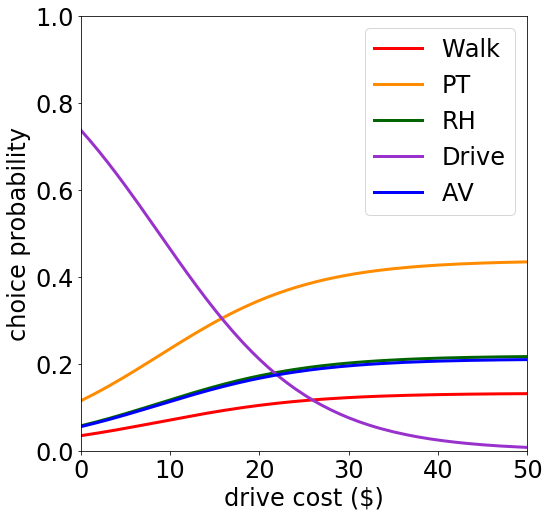}\label{sfig:substitution_mnl_8k_sgp}}\\
\subfloat[Opt-DNNs (8K-LD)]{\includegraphics[width=0.23\linewidth]{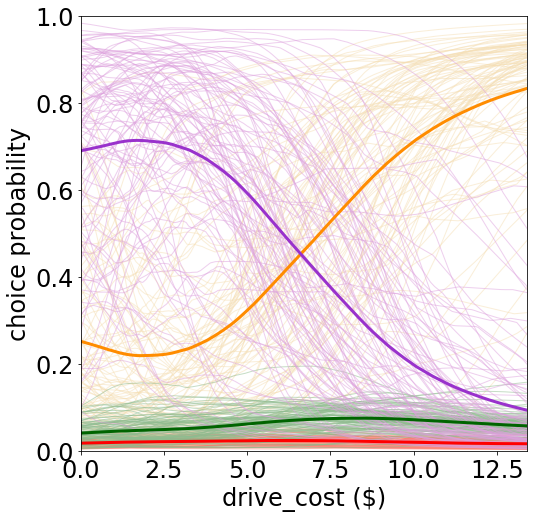}\label{sfig:substitution_opt_8k_ld}}
\subfloat[Opt-DNNs (80K-LD)]{\includegraphics[width=0.23\linewidth]{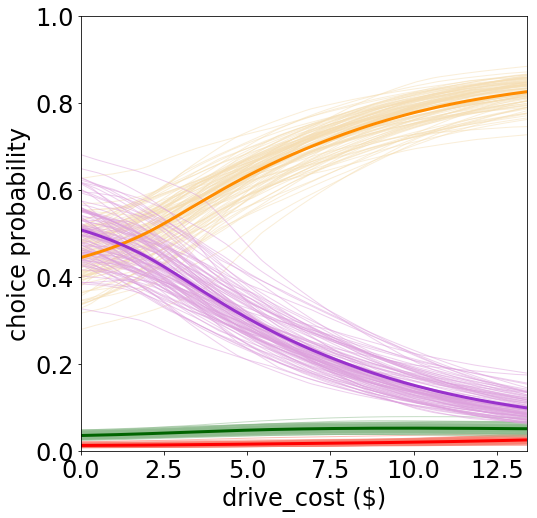}\label{sfig:substitution_opt_80k_ld}}
\subfloat[Random-DNNs (80K-LD)]{\includegraphics[width=0.23\linewidth]{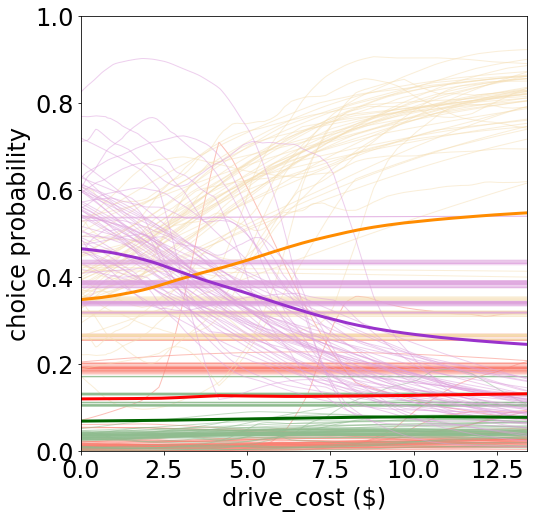}\label{sfig:substitution_random_80k_ld}}
\subfloat[MNL (80K-LD)]{\includegraphics[width=0.23\linewidth]{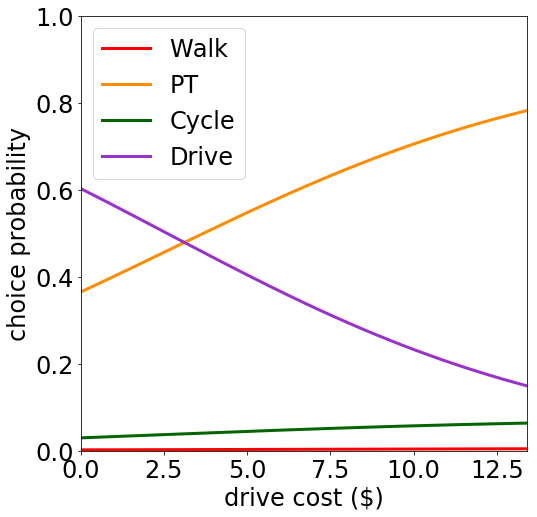}\label{sfig:substitution_mnl_80k_ld}}\\
\caption{Substitution patterns of all the alternatives with varying driving costs}
\label{fig:substitution}
\end{figure}

\noindent
The substitution pattern of the alternatives is of both practical and theoretical importance in choice analysis. In practice, researchers need to understand how market shares vary with input variables; in theory, the substitution pattern constitutes the critical difference between multinomial logit, nested logit, and mixed logit models. Figure \ref{fig:substitution} visualizes how the choice probability functions vary as the driving cost increases. By visualizing the choice probabilities of all the alternatives, Figure \ref{fig:substitution} is an one-step extension of Figure \ref{fig:prob}.

The substitution pattern of the Opt-DNNs is more reasonable than that of the Random-DNNs and more flexible than that of the MNL models. As shown in Figure \ref{sfig:substitution_opt_8k_sgp}, when the driving cost is smaller than $\$20$, the substitution pattern of the Opt-DNNs aggregated over the $100$ models illustrates that the five alternatives are substitute to each other, since the driving probability is decreasing while others are increasing. When the driving cost is larger than $\$20$, the substitution pattern between walking, ridesharing, driving, and using an AV still reveals the substitute nature. This reasonable substitution pattern is retained in the LD dataset (Figures \ref{sfig:substitution_opt_8k_ld} and \ref{sfig:substitution_opt_80k_ld}). In a choice modeling setting, the alternatives in a choice set are typically substitutes: people are expected to switch from driving to other travel modes, as the driving cost increases. Therefore, the aggregated substitution pattern has mostly reflected the correct relationship of the five alternatives. 

However, the three theoretical challenges also permeate into the substitution patterns, particularly when the sample size is small. The large variation in Figures \ref{sfig:substitution_random_8k_sgp} and \ref{sfig:substitution_random_80k_ld} illustrates the high sensitivity to hyperparameters; the large variation in Figures \ref{sfig:substitution_opt_8k_sgp}, \ref{sfig:substitution_opt_8k_ld}, and \ref{sfig:substitution_opt_80k_ld} illustrates the model non-identification problem; and the individual curves in Figures \ref{sfig:substitution_opt_8k_sgp}, \ref{sfig:substitution_opt_8k_ld} reveal the local irregularity. Note that larger sample, hyperparameter searching, and regularization methods can mitigate but not fully solve these problems. For example in Figure \ref{sfig:substitution_opt_8k_sgp}, the average substitution pattern of the Opt-DNNs indicate that people are less likely to choose the public transit as the driving cost increases, when the driving cost is larger than $\$20$. As a comparison, the substitution pattern in the two MNL models, although perhaps exceedingly restrictive, reflects the travel mode alternatives being substitute goods. 

\subsubsection{Market Shares}
\label{sec:ms}
\noindent
Table \ref{table:market_share} summarizes the market shares predicted by the three model groups in the 8K-SGP and 80K-LD datasets. We found that, while the choice probability functions of Opt-DNNs can be locally unreasonable, the aggregated market shares of Opt-DNNs are very close to the true market shares, and the market shares predicted by Opt-DNNs are as accurate as the MNL in the testing sets. Specifically, in both the training and testing sets of the 8K-SGP and 80K-LD data sets, the errors between the predicted market shares of Opt-DNNs and the true market shares are within the range of $1.0\%$. It appears that the three challenges of DNNs do not emerge in the calculation of market shares. The local irregularity could be cancelled out owing to the aggregation over the sample. The model non-identification appears less a problem as the market shares across the Opt-DNN trainings are very stable, as shown by the small standard deviations in the parenthesis. The high sensitivity to hyperparameters is addressed by the selection of the Opt-DNNs from the hyperparameter searching process, as the market shares of the Opt-DNNs are much more accurate than the Random-DNNs. It is also interesting to compare Opt-DNNs and MNL: while MNL is guaranteed to provide market shares exactly the same as the true ones in the training sets \footnote{The first order conditions in training MNL actually match the estimated and true market shares in the training sets.}, the predicted market shares of Opt-DNNs are as accurate as MNL in the testing sets. Specifically, we computed the sum of the absolute errors between the predicted and the true market shares for Opt-DNNs and MNL in the testing sets. In the 8K-SGP, the absolute error of Opt-DNNs is $4.28\%$, slightly higher than $3.05\%$ from MNL; in the 80K-LD, the absolute error of Opt-DNNs is $1.90\%$, slightly lower than $2.10\%$ from MNL.

\begin{table}[htb]
\centering
\caption{Market shares of travel modes (training and testing); each entry represents the average value of the market share over $100$ trainings, and the number in the parenthesis is the standard deviation. }
\resizebox{1.0\linewidth}{!}{%
\begin{tabular}{p{0.2\linewidth} | P{0.25\linewidth} P{0.25\linewidth} P{0.25\linewidth} | P{0.2\linewidth}}
        \toprule
        \hline
        \multicolumn{5}{l}{\textbf{Panel 1. Training sets}} \\
        \hline
        & Opt-DNNs (8K-SGP) & Random-DNNs (8K-SGP) & MNL (8K-SGP) & True Market Share \\
        \hline
        Walk & 10.2\% (0.6\%) &  12.7\% (2.4\%)& 10.6\% & 10.6\% \\
        Public Transit & 23.1\% (1.0\%) &20.8\% (2.8\%)  & 23.0\% & 23.0\% \\
        Ride Hail & 10.5\% (0.6\%) & 12.7\% (2.4\%) & 10.7\% & 10.7\% \\
        Drive & 45.6\% (0.8\%) & 41.0\% (4.5\%) & 44.9\% & 44.9\% \\
        AV & 10.6\% (0.6\%) & 12.80\% (2.4\%) & 10.8\% & 10.8\% \\
        \hline
        & Opt-DNNs (80K-LD) & Random-DNNs (80K-LD) & MNL (80K-LD) & True Market Share \\
        \hline
        Walk & 17.9\% (1.8\%) & 19.7\% (3.6\%) & 17.6\% & 17.6\% \\
        Public Transit & 35.1\% (2.5\%) & 32.7\% (3.8\%) & 35.3\% & 35.3\% \\
        Cycle & 2.9\% (0.3\%) & 6.9\% (4.6\%) & 3.0\% & 3.0\% \\
        Drive & 44.1\% (2.8 \%) & 40.7\% (4.6\%) & 44.1\% & 44.1\% \\
        \midrule
        \hline
        \multicolumn{5}{l}{\textbf{Panel 2. Testing sets}} \\
        \hline
        & Opt-DNNs (8K-SGP) & Random-DNNs (8K-SGP) & MNL (8K-SGP) & True Market Share \\
        \hline
        Walk & 9.1\% (1.3\%) & 12.3\% (2.7\%) & 10.34\% & 9.48\% \\
        Public Transit & 23.4\% (2.1\%) & 21.0\% (3.2\%) & 23.1\% & 23.9\% \\
        Ride Hail & 10.3\% (1.2\%) & 12.7\% (2.5\%) & 10.5\% & 10.8\% \\
        Drive & 46.7\% (1.8\%) & 41.2\% (4.9\%) & 45.2\% & 44.5\% \\
        AV & 10.5\% (1.3\%) & 12.8\% (2.5\%) & 10.8\% & 11.2\% \\
        \hline
        & Opt-DNNs (80K-LD) & Random-DNNs (80K-LD) & MNL (80K-LD) & True Market Share \\
        \hline
        Walk & 18.0\% (1.8\%) & 19.8\% (3.6\%) & 17.7\% & 17.3\% \\
        Public Transit & 35.0\% (2.5\%) & 32.6\% (3.8\%) & 35.3\% & 34.7\% \\
        Cycle & 2.8\% (0.3\%) & 6.9\% (4.6\%) & 2.9\% & 2.8\% \\
        Drive & 44.2\% (2.8\%) & 40.7\% (4.6\%) & 44.1\% & 45.1\% \\        
        \bottomrule
    \end{tabular}
} 
\label{table:market_share}
\end{table}

\subsubsection{Social Welfare}
\label{sec:swa}
\noindent
Since DNNs have an implicit utility interpretation, we can observe how social welfare changes as action variables change the values. To demonstrate this process, we simulate one dollar decrease of the driving cost, and calculate the average social welfare change in the Opt-DNNs in the 8K-SGP dataset. We found that the social welfare increases by about $\$520$ in the Opt-DNN models after averaging over all $100$ trainings. Interestingly, the magnitude of this social welfare change ($\$520$) is very intuitive and consistent with the one computed from MNL models, which is $\$491$ dollars. In the process of computing the social welfare change, we used the $\alpha_i$ averaged across $100$ trainings as the individual $i$'s marginal value of utility, slightly different from the formula in Table \ref{table:dnn_econ_info}. Specifically, the formula is
$\sum_i \frac{1}{\bar{\alpha_i}} \big[ \log (\sum_{k=1}^K e^{\hat{V}^1_{ik}}) - \log (\sum_{k=1}^K e^{\hat{V}^0_{ik}}) \big]$
, in which $\bar{\alpha_i} = \frac{1}{S} \sum_s \alpha_{i,s}$; $\alpha_{i,s}$ represents individual $i$'s marginal value of utility for each DNN model $s$; ${V}^1_{ik}$ and ${V}^0_{ik}$ represent the utility values under two different scenarios. Without using $\bar{\alpha_i}$, individuals' marginal value of utility can take unreasonable values. The problem associated with the disaggregate gradient information will be discussed in details in the following section.

\subsection{Gradient-Based Interpretation}
\subsubsection{Gradients of Choice Probability Functions}
\label{sec:gcpf}
\noindent
The gradient of choice probability functions offers opportunities to extract more important economic information. Since researchers often seek to understand how to take actions to trigger behavioral changes, the most relevant information is the partial derivative of the choice probability function with respect to a targeting input variable. Figure \ref{fig:prob_derivative} visualizes the corresponding derivatives of the choice probability functions. As shown below, both the strength and the challenges identified in the choice probability functions are retained in the probability derivatives.

\begin{figure}[ht]
\captionsetup[subfigure]{justification=centering}
\centering
\subfloat[Opt-DNNs (8K-SGP)]{\includegraphics[width=0.23\linewidth]{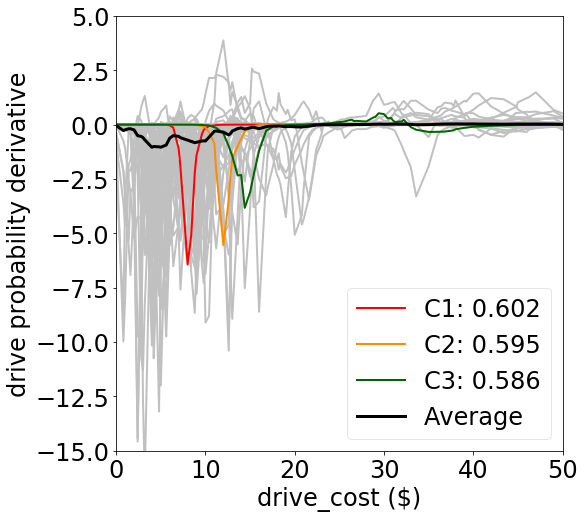}\label{sfig:prob_derivative_opt_8k_sgp}}
\subfloat[Random-DNNs (8K-SGP)]{\includegraphics[width=0.23\linewidth]{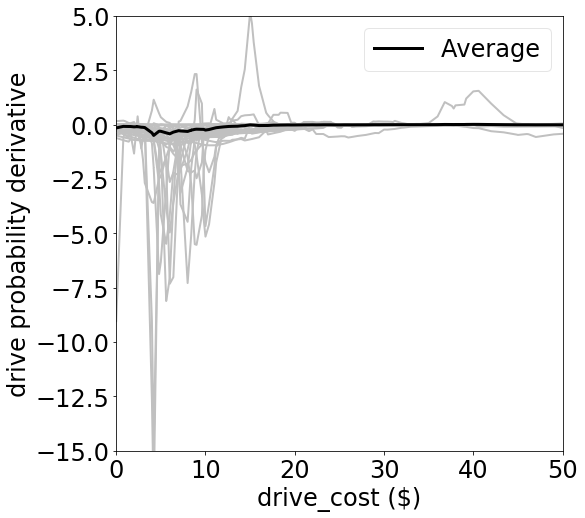}\label{sfig:prob_derivative_random_8k_sgp}}
\subfloat[MNL (8K-SGP)]{\includegraphics[width=0.23\linewidth]{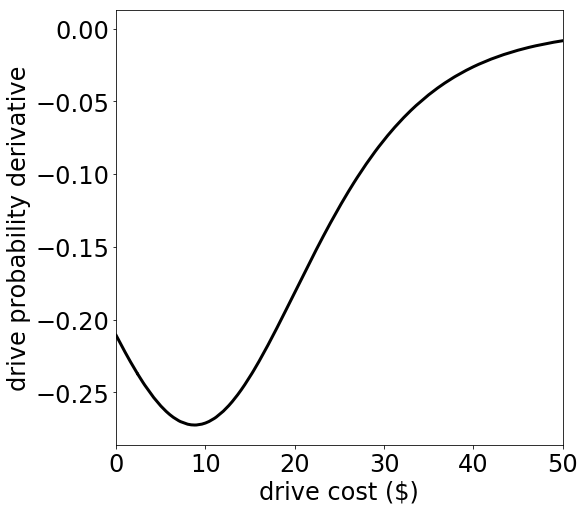}\label{sfig:prob_derivative_mnl_8k_sgp}}\\
\subfloat[Opt-DNNs (8K-LD)]{\includegraphics[width=0.23\linewidth]{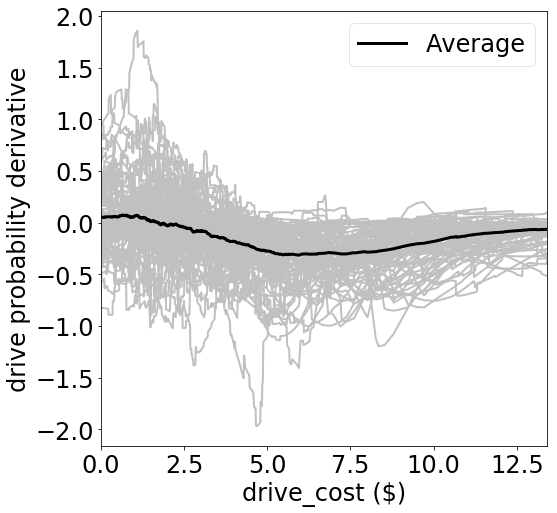}\label{sfig:prob_derivative_opt_8k_ld}}
\subfloat[Opt-DNNs (80K-LD)]{\includegraphics[width=0.23\linewidth]{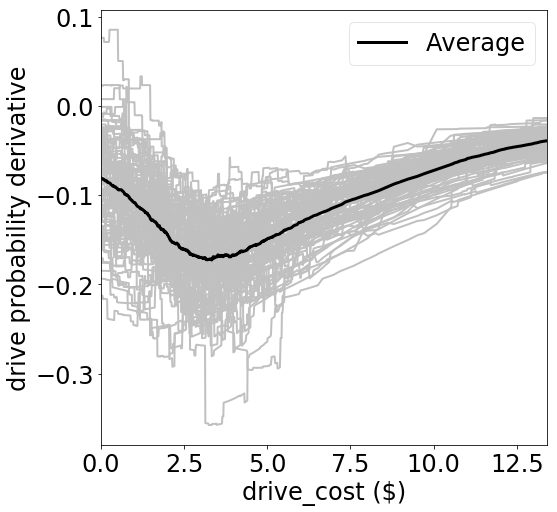}\label{sfig:prob_derivative_opt_80k_ld}}
\subfloat[Random-DNNs (80K-LD)]{\includegraphics[width=0.24\linewidth]{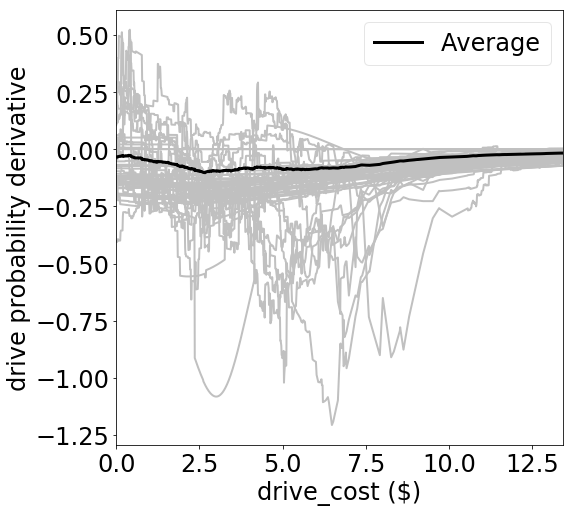}\label{sfig:prob_derivative_random_80k_ld}}
\subfloat[MNL (80K-LD)]{\includegraphics[width=0.25\linewidth]{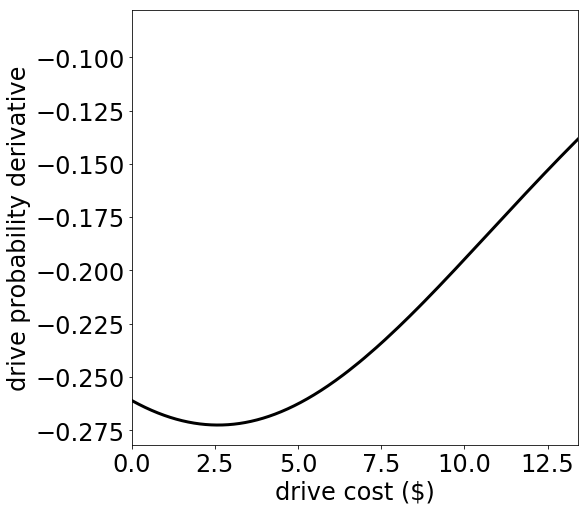}\label{sfig:prob_derivative_mnl_80k_ld}}\\
\caption{Probability derivatives of choosing driving with varying driving costs}
\label{fig:prob_derivative}
\end{figure}

In Figure \ref{sfig:prob_derivative_opt_8k_sgp}, the majority of the Opt-DNNs, such as the three curves (C1, C2, and C3), take negative values and have inverse bell shapes. This inverse bell shaped curve is intuitive because people are not as sensitive to price changes when price is close to zero or infinity, but are more sensitive when price is close to a certain tipping point. The shapes revealed by Opt-DNNs are similar to the MNL models. The probability derivative of MNL models is $\partial s(x) / \partial x = s(x) (1 - s(x)) \times (\partial V(x) / \partial x)$, which is mostly negative and take a very regular truncated inverse bell shape, as shown in Figures \ref{sfig:prob_derivative_mnl_8k_sgp} and \ref{sfig:prob_derivative_mnl_80k_ld}. 

The sensitivity to hyperparameters, the model non-identification, and the local irregularity also exist here for probability derivatives. Random-DNNs reveal more unreasonable behavioral patterns than Opt-DNNs, as many of the input gradients are flat on zero, demonstrating the importance of selecting correct hyperparameters. The variation of individual trainings in Figures \ref{sfig:prob_derivative_opt_8k_sgp}, \ref{sfig:prob_derivative_opt_8k_ld}, and \ref{sfig:prob_derivative_opt_80k_ld} demonstrates the challenge of model non-identification. With fixed training samples and hyperparameters, the DNN trainings can lead to different training results, thus creating difficulty for researchers to choose a final model for interpretation. The exploding gradients and the non-monotonicity issues, as the two indicators of local irregularity, are also clearly illustrated in the individual trainings in Figures \ref{sfig:prob_derivative_opt_8k_sgp} and \ref{sfig:prob_derivative_opt_8k_ld}, although are less severe in Figure \ref{sfig:prob_derivative_opt_80k_ld}. The absolute values of many probability derivatives are of large magnitude; for example in Figure \ref{sfig:prob_derivative_opt_8k_sgp}, at the peak of the C1 curve, $\$1$ increase of driving costs leads to about $6.5\%$ change in choice probability\footnote{This $6.5\%$ appears much smaller than the values in Figure \ref{fig:prob}. It is because of the difference between arc and point elasticities.}, which is much larger than the MNL models. Similar to the previous discussions, large sample size, hyperparameter searching, model ensemble, regularization, and information aggregation can mitigate these challenges.

\subsubsection{Elasticities}
\noindent
To compare across input variables, researchers often compute elasticities because the elasticities are standardized derivatives. Given that DNNs provide choice probability derivatives, it is straightforward to compute the elasticities from DNNs. For MNL models, the formula to compute elasticities are attached in Appendix IV. Tables \ref{table:elasticities_dnn} and \ref{table:elasticities_80k_ld} present the elasticities of travel mode choices with respect to input variables. In Panel 1, each entry represents the average elasticity of the respondents in the testing set based on one Opt-DNN model, and the value in the parenthesis is the standard deviation of individuals' elasticity values. Panel 2 is the average elasticity of the testing set from a MNL model with linear specification, and the value in the parenthesis represents the standard deviation.

\begin{table}[htb]
\centering
\caption{Elasticities of five travel modes with respect to input variables (8K-SGP dataset)}
\resizebox{1.0\linewidth}{!}{%
\begin{tabular}{p{0.35\linewidth} P{0.15\linewidth} P{0.15\linewidth} P{0.15\linewidth} P{0.15\linewidth} P{0.15\linewidth}}
\toprule
        \textbf{Panel 1: DNN Model} & Walk & Public Transit & Ride Hailing & Driving & AV \\
        \midrule
        Walk time & \textbf{-5.308(6.9)} & 0.399(5.9) & -0.119(7.1) & -0.030(4.6) & -1.360(6.8) \\
        Public transit cost & -1.585(9.6) & \textbf{-4.336(9.6)} & -1.648(11.1) & 1.081(5.9) & 1.292(9.5) \\
        Public transit walk time & 0.123(6.9) & \textbf{-1.707(6.5)} & 0.047(7.3) & 0.621(4.7) & 0.844(6.7) \\
		public transit wait time & 0.985(8.7) & \textbf{-2.520(8.9)} & -0.518(9.1) & 0.092(5.8) & 0.366(8.8)\\
		Public transit in-vehicle time & 0.057(9.0) & \textbf{-1.608(9.0)} & 0.484(9.4) & 0.778(5.8) & 1.273(8.9)\\
        Ride hail cost & -2.353(7.6) & 0.005(6.9) & \textbf{-4.498(8.9)} & 0.304(5.6) & -0.243(9.0) \\
        Ride hail wait time & 0.234(8.8) & 1.471(8.3) & \textbf{-2.536(10.1)} & -0.253(5.7) & -0.228(8.8) \\
		Ride hail in-vehicle time & 0.299(7.8) & -0.224(7.4) & \textbf{-5.890(9.4)} & 0.740(5.4) & 0.739(7.6) \\
        Drive cost & 1.124(6.6) & 2.545(5.9) & 3.760(6.8) & \textbf{-1.886(5.0)} & 2.273(6.9)\\
        Drive walk time &  2.033(5.3) & 0.552(5.0) & 2.503(5.6) & \textbf{-0.412(3.8)} & 1.787(5.4)\\
		Drive in-vehicle time & 1.824(9.0) & 4.163(8.2) & 3.640(9.9) & \textbf{-3.199(7.4)} & 3.268(9.1)\\
        AV cost & -0.562(6.5) & -0.198(6.2) & 0.819(6.9) & 0.337(4.6) & \textbf{-4.289(7.6)} \\
        AV wait time & -0.068(7.9) & -0.695(7.4) & 2.400(8.4) & 0.284(4.6) & \textbf{-1.591(7.8)} \\
		AV in-vehicle time & -0.784(6.2) & 0.221(5.6) & 0.955(7.1) & 0.079(4.3) & \textbf{-4.534(6.8)}\\
		Age & -1.003(18.7) & 2.502(18.4) & -4.385(20.0) & 0.949(13.7) & -1.936(18.6)\\
		Income & 1.127(10.7) & 0.727(10.5) & 0.957(11.9) & -0.002(6.7) & 2.539(10.8)\\
        \hline
        \hline        
        \textbf{Panel 2: MNL Model} & Walk & Public Transit & Ride Hailing & Driving & AV \\
        \midrule
        Walk time & \textbf{-1.916(1.8)} & 0.130(0.1) & 0.130(0.1) & 0.130(0.1) & 0.130(0.1) \\ 
        Public transit cost & 0.137(0.1) & \textbf{-0.566(0.4)} & 0.137(0.1) & 0.137(0.1) & 0.137(0.1) \\ 
        Public transit access time & 0.083(0.1) & \textbf{-0.318(0.3)} & 0.083(0.1) & 0.083(0.1) & 0.083(0.1) \\ 
        Public transit transfer time & 0.072(0.1) & \textbf{-0.265(0.2)} & 0.072(0.1) & 0.072(0.1) & 0.072(0.1) \\ 
        Public transit in-vehicle time & 0.126(0.1) & \textbf{-0.478(0.4)} & 0.126(0.1) & 0.126(0.1) & 0.126(0.1) \\ 
        Ride hail cost & 0.028(0.0) & 0.028(0.0) & \textbf{-0.248(0.2)} & 0.028(0.0) & 0.028(0.0) \\ 
        Ride hail wait time & 0.033(0.0) & 0.033(0.0) & \textbf{-0.304(0.2)} & 0.033(0.0) & 0.033(0.0) \\ 
        Ride hail in-vehicle time & 0.076(0.1) & 0.076(0.1) & \textbf{-0.716(0.5)} & 0.076(0.1) & 0.076(0.1) \\ 
        Drive cost & 0.292(0.2) & 0.292(0.2) & 0.292(0.2) & \textbf{-0.756(1.1)} & 0.292(0.2) \\ 
        Drive walk time & 0.120(0.1) & 0.120(0.1) & 0.120(0.1) & \textbf{-0.211(0.3)} & 0.120(0.1) \\ 
        Drive in-vehicle time & 0.291(0.2) & 0.291(0.2) & 0.291(0.2) & \textbf{-0.463(0.6)} & 0.291(0.2) \\ 
        AV cost & 0.044(0.1) & 0.044(0.1) & 0.044(0.1) & 0.044(0.1) & \textbf{-0.413(0.4)} \\ 
        AV wait time & 0.029(0.0) & 0.029(0.0) & 0.029(0.0) & 0.029(0.0) & \textbf{-0.254(0.2)} \\ 
        AV in-vehicle time & 0.067(0.1) & 0.067(0.1) & 0.067(0.1) & 0.067(0.1) & \textbf{-0.638(0.6)} \\ 
        Age & -0.168(0.1) & 0.546(0.2) & -0.695(0.2) & 0.002(0.1) & -0.363(0.2) \\ 
        Income & -0.102(0.1) & -0.177(0.1) & 0.061(0.0) & 0.056(0.0) & 0.148(0.1) \\ 
    \bottomrule
    \end{tabular}
} 
\label{table:elasticities_dnn}
\end{table}

The average elasticities in the Opt-DNN are reasonable in terms of both the signs and magnitudes. We highlight the elasticities that relate the travel modes to their own alternative-specific variables. These highlighted elasticities are all negative, which is very reasonable since higher travel cost and time should lead to lower probability of adopting the corresponding travel mode. In Table \ref{table:elasticities_dnn}, the magnitudes in the DNN models are higher than the typical results from the MNL models, although the relative magnitudes of the elasticity coefficients in DNNs are similar to the MNL model. For example, DNN models indicate that $1\%$ increase in public transit cost, walking time, waiting time, and in-vehicle travel time leads to the decrease of $4.3 \%$, $1.7 \%$, $2.5 \%$, and $1.6 \%$ probabilities in using public transit, and the absolute magnitudes of these numbers are larger than but the relative magnitudes are similar to the MNL model, in which the corresponding probability decreases are $0.56 \%$, $0.31 \%$, $0.26 \%$, and $0.48 \%$. This difference in the absolute magnitude is already manifested in the previous discussion that the gradients of DNN models are larger than that of MNL models. In addition, the highlighted self-elasticities in the DNNs are overall of a larger magnitude than the cross-elasticity values, which is also reasonable. 

Local irregularity is revealed here by the large standard deviations of the elasticities. For example in Table \ref{table:elasticities_dnn}, as the walking elasticity regarding walking time is $-5.3$ on average, its standard deviation is $6.9$. This large standard deviation is caused by local irregularity, as individuals can have dramatically different elasticity values. The other two challenges, the high sensitivity and the model non-identification, are not presented in the process of computing the average elasticities, because the Opt-DNNs are trained by the same set of hyperparameters and the model non-identification is not seen in only one Opt-DNN.

\begin{table}[htb]
\centering
\caption{Elasticities of travel modes with respect to input variables (80K-LD dataset)}
\resizebox{0.85\linewidth}{!}{%
    \begin{tabular}{p{0.35\linewidth} | P{0.15\linewidth} P{0.15\linewidth} P{0.15\linewidth} P{0.15\linewidth}}
    \toprule
        \textbf{Panel 1: DNN Model} & Walk & Public Transit & Cycle & Driving \\
        \midrule
        Walk time & \textbf{-1.494(0.8)} & -0.437(0.9) & -0.926(1.0) & 0.680(0.7) \\
        Public transit cost & 0.260(0.3) & \textbf{-0.199(0.3)} & 0.118(0.3) & -0.002(0.2) \\
        Public transit access time & 0.310(0.5) & \textbf{-0.590(0.6)} & 0.012(0.4) & 0.239(0.3)  \\
    	Public transit transfer time & -0.028(0.2) & \textbf{-0.118(0.3)} & -0.073(0.2) & 0.069(0.2)\\
    	Public transit in-vehicle time & 0.128(0.3) & \textbf{-0.324(0.6)} & 0.026(0.4) & 0.217(0.4) \\
        Cycle time &  -0.134(0.2) & 0.376(0.4) & \textbf{0.113(0.3)} & -0.167(0.3) \\
        Drive cost & 0.070(0.2) & 0.039(0.1) & 0.025(0.1) & \textbf{-0.127(0.3)} \\
    	Drive in-vehicle time & 0.280(0.5) & 1.047(1.0) & 0.561(0.8) & \textbf{-0.840(0.9)} \\
        Age & -0.236(0.7) & -0.008(0.8) & -0.223(0.8) & 0.204(0.8)\\ 
        \midrule
        \textbf{Panel 2: MNL Model} & Walk & Public Transit & Cycle & Driving \\
        \midrule
        Walk time & \textbf{-7.344(7.8)} & 0.334(0.4) & 0.334(0.4) & 0.334(0.4) \\ 
        Public transit cost & 0.049(0.1) & \textbf{-0.066(0.1)} & 0.049(0.1) & 0.049(0.1) \\ 
        Public transit access time & 0.262(0.3) & \textbf{-0.459(0.4)} & 0.262(0.3) & 0.262(0.3) \\ 
        Public transit transfer time & 0.093(0.2) & \textbf{-0.110(0.2)} & 0.093(0.2) & 0.093(0.2) \\ 
        Public transit in-vehicle time & 0.222(0.3) & \textbf{-0.261(0.3)} & 0.222(0.3) & 0.222(0.3) \\ 
        Cycle time & 0.019(0.0) & 0.019(0.0) & \textbf{-0.757(0.7)} & 0.019(0.0) \\ 
        Drive cost & 0.070(0.1) & 0.070(0.1) & 0.070(0.1) & \textbf{-0.235(0.5)} \\ 
        Drive in-vehicle time & 0.811(0.9) & 0.811(0.9) & 0.811(0.9) & \textbf{-1.269(1.7)} \\ 
        Age & -0.275(0.2) & 0.163(0.1) & -0.304(0.2) & -0.004(0.1) \\ 
    \bottomrule
    \end{tabular}
} 
\label{table:elasticities_80k_ld}
\end{table}

The difference between the Opt-DNN and the MNL in terms of the coefficient magnitude seems less salient in the 80K-LD dataset, as shown in Table \ref{table:elasticities_80k_ld}. Note that the coefficients of the DNN model on the main diagonal in Panel 1 are mainly negative, which are the same as the findings in Table \ref{table:elasticities_dnn}. Interestingly, even the absolute magnitudes of the DNN models become similar to the MNL models. This result reflects the smoother choice probability functions in the large sample compared to the small sample cases, as discussed in Figure \ref{fig:prob}. It would be very difficult to provide a definitive answer to the question what size can be treated as ``large'' for DNN models. We will discuss this in the last section, but here we can conclude that the average elasticity coefficients from DNNs are largely intuitive.

\subsubsection{Marginal Rates of Substitution: Heterogeneous Values of Time}
\label{sec:mrs}
\begin{figure}[ht]
\captionsetup[subfigure]{justification=centering}
\centering
\subfloat[Opt-DNNs (8K-SGP)]{\includegraphics[width=0.3\linewidth]{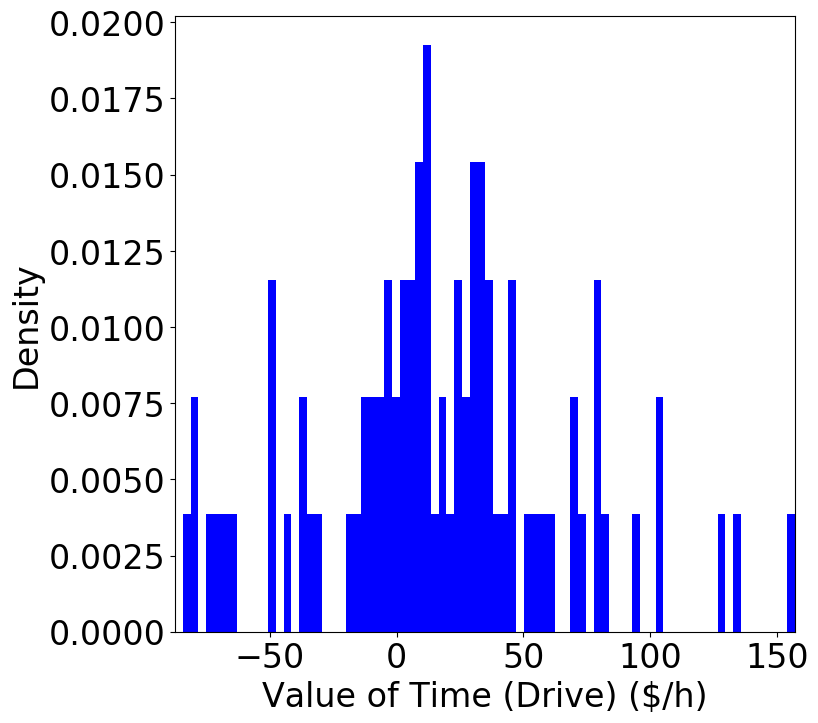}\label{sfig:vot_opt_8k_sgp}}
\subfloat[Opt-DNNs (80K-LD)]{\includegraphics[width=0.295\linewidth]{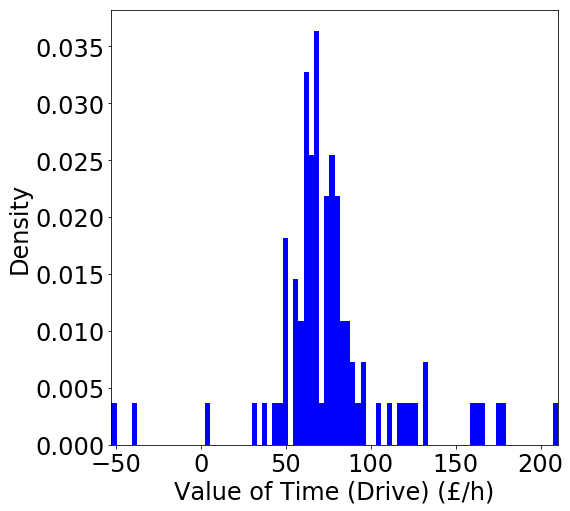}\label{sfig:vot_opt_80k_ld}} \\
\subfloat[One Opt-DNN (8K-SGP)]{\includegraphics[width=0.3\linewidth]{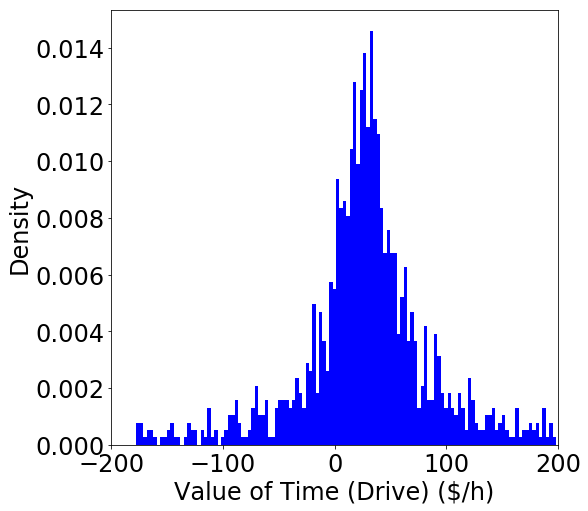}\label{sfig:vot_one_opt_8k_sgp}}
\subfloat[One Opt-DNN (80K-LD)]{\includegraphics[width=0.295\linewidth]{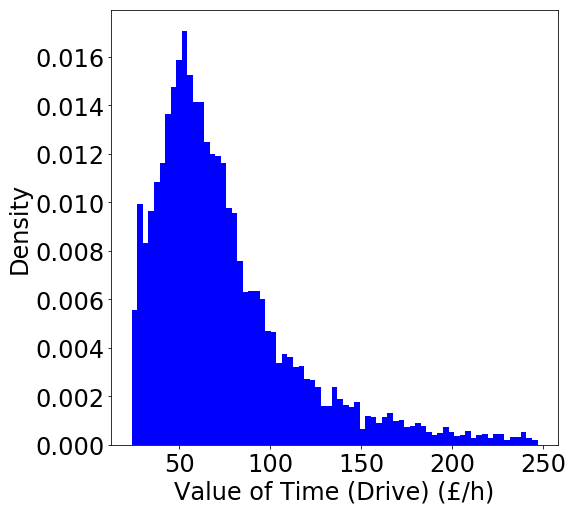}\label{sfig:vot_one_opt_80k_ld}} 
\caption{Heterogeneous values of time; the extremely large and small values (below $5\%$ percentile and above $95\%$ percentile) are cut off from this histogram.}
\label{fig:vot}
\end{figure}

\noindent
VOT, as one example of MRS, is one of the most important pieces of economic information obtained from choice models, since the monetary gain from time saving is the most prevalent benefit from the improvement of any transportation system. As VOT is computed as the ratio of two parameters in a MNL model, the ratio of two probability derivatives represents the VOT in DNNs. Figures \ref{sfig:vot_opt_8k_sgp} and \ref{sfig:vot_opt_80k_ld} represent the distribution of the average VOT among all individuals over 100 trainings of Opt-DNNs; Figures \ref{sfig:vot_one_opt_8k_sgp} and \ref{sfig:vot_one_opt_80k_ld} represent the distribution of the heterogeneous VOT of the individuals in one Opt-DNN model. The distributions of VOT in Figures \ref{sfig:vot_one_opt_8k_sgp} and \ref{sfig:vot_one_opt_80k_ld} resemble the typical analysis about the heterogeneous VOT in DCMs.

The aggregate information such as the average values of the VOT distributions are quite reasonable, while certain regions of the VOT distributions can be somewhat counter-intuitive as they may have a very large dispersion and even some negative values. For example, the median VOT in the testing set of the 8K-SGP dataset is $\$27.8/h$, and the VOT distribution is highly concentrated around its mean value, resembling the shape of a Gaussian distribution. Similar patterns can be observed for the 80K-LD dataset in Figures \ref{sfig:vot_one_opt_80k_ld}. This finding of reasonable aggregate information but irregular disaggregate information is consistent with our previous findings. 

The median $\$27/h$ VOT in Figure \ref{sfig:vot_opt_8k_sgp} is consistent with previous studies, in which VOT has been found to be between $\$7.8/h$ and $\$30.3/h$ for various travel modes \cite{Ho2016}. VOT has also been found to be between $21\%$ and $254\%$ of the hourly wage in a review paper \cite{Zamparini2016}. By using the average hourly wage ($\$27.16/h$) of the U.S workers in 2018, we would expect the VOT to be between $\$5.7/h$ and $\$70.0/h$. Our VOT obtained from DNNs is about in the middle of this range. Intuitively, the VOT should be of the same magnitude as the hourly wages, and $\$27/h$ is very close to the average hourly wage. However, on the other hand, the VOT obtained from DNNs can be unreasonable for certain individuals. It is highly unlikely for VOT to be negative, while DNNs detect a sizeable portion of people whose VOT are negative. This counter-intuitive result is caused by the local irregularity of the probability derivatives. But on the other side, larger sample size can mitigate this problem even at the individual level. In the 80K-LD dataset, the range of VOT is mainly between $\$40.0/h$ and $\$80.0/h$ and the median value is $\$69.1/h$. Both the range and the median value are reasonable based on the findings of the previous studies.

\section{Conclusions and Discussions}
\label{s:5}
\noindent
This study aims to interpret DNN models in the context of choice analysis and extract economic information as complete as obtained from classical DCMs. The economic information includes a complete list of choice predictions, choice probabilities, market share, substitution patterns of alternatives, social welfare, probability derivatives, elasticity, marginal rates of substitution (MRS), and heterogeneous values of time (VOT). The process of interpreting DNN models is different from classical DCMs because DNNs are a very flexible model family, capable of automatically learning more flexible behavioral patterns than the regular patterns pre-specified by domain experts in the classical DCMs. As a result, we found that most economic information extracted from DNNs is reasonable and more flexible than the MNL models, particularly when the sample size is large. However, the economic information automatically learnt by DNNs can be sometimes unreliable, caused by three challenges: high sensitivity to hyperparameters, model non-identification, and local irregularity. Owing to the high sensitivity to hyperparameters, the DNN models without appropriate regularizations or architectures cannot provide valuable economic information. Owing to the model non-identification, researchers cannot obtain an ultimate function estimate for economic interpretation. Owing to the local irregularity, DNN models reveal unreasonable local behavioral patterns when the sample size is small. These three problems can be partially addressed by using large sample, simple random hyperparameter searching, model ensemble, regularization, and information aggregation. Particularly, the economic information based on the ensemble method, such as the average choice probability function, average probability derivatives, market shares, average social welfare change, average elasticities, and the median VOT, are mostly consistent with our behavioral intuition and previous studies.

There should be little doubt that DNNs can provide a full set of economic information as DCMs; however, many questions remain. As shown in this study, the 80K-LD dataset can provide more reasonable behavioral patterns than the 8K-LD and 8K-SGP datasets, demonstrating the importance of a large sample size. One challenging question is the exact sample size that can be counted as ``large'' for DNNs. Unfortunately, it is very difficult to provide a definitive answer, since it always depends on other factors such as model complexity and input dimensions. In principle, as models become more complicated, researchers need larger sample size, as shown in the proof about the estimation error of DNNs in Appendix I. The exact model complexity of DNN models empirically depends on the specific DNN architectures and is also theoretically ambiguous owing to the difficulty of deriving a tight upper bound on the estimation errors. The bottom line, regarding the DNN-based choice models, is that the sample size traditionally counted as large, such as thousands of observations, seem inadequate to provide reliable economic information in DNN-based choice models. This is not a surprise since the model complexity of DNNs is much larger than classical DCMs.  While our study suggests that about $O(10^4)$ sample size seems to be adequate for economic information, we would encourage future studies to further explore this question, particularly when more complicated models and inputs, such as images and natural languages, are involved in choice modeling settings.

We also discussed the irregular behavioral patterns of the utility functions in the 8K-LD and 8K-SGP datasets. However, we would like to emphasize that this ``irregularity'' does not necessarily have a negative connotation. For example, while certain patterns in DNNs can be treated as unreasonable from the perspective of classical MNL models, these patterns exist in other types of behavioral and machine learning literature. As to the exploding gradients, behavioral theory suggests that people can have a sharp threshold price in decision-making, and as a result, the gradients around the threshold price can be very large. The decision tree model suggests a similar behavioral mechanism: the input gradients around the cutoff points can take very large values. As to the non-monotonicity pattern, people might become more likely to buy certain commodity when the price of the commodity increases, since people treat higher prices as the signal of better quality of the commodities, leading to positive elasticity values of choice probability functions regarding the price variables. In short, the patterns revealed in the 8K-SGP and 8K-LD datasets can be positively evaluated as a success for identifying certain flexible behavioral patterns that cannot be found in a restrictive MNL model, or as a failure that identifies unreasonable and unrealistic behavioral patterns caused by small sample and high model complexity. We do not take a stance here, but leave this question for future studies.

This study has demonstrated the importance of using hyperparameter searching, repeated trainings, regularization methods, and aggregation over models and population to improve the reliability of the economic information. However, it remains an open question what the most effective methods are in terms of making the economic information more reliable from the classical DCM perspective. Recent studies in the ML community have suggested potential methods to address the three challenges. As to the high sensitivity to hyperparameters, potential remedies include a large number of regularization methods, such as domain constraints, Bayesian priors, model ensemble \cite{Krizhevsky2012}, data augmentation \cite{Bishop1995}, dropouts \cite{Hinton2012}, early stopping, and sparse connectivity; new DNN architectures, such as AlexNet \cite{Krizhevsky2012}, GoogleNet \cite{Szegedy2015}, and ResNet \cite{HeKaiming2016}; or smarter ways to tune hyperparameters, such as construction of a continuous hyperparameter space, Gaussian process, Bayesian neural networks \cite{Snoek2012,Snoek2015}, or reinforcement learning \cite{Zoph2016,Zoph2017,Baker2016}, much richer than a simple random searching in discrete grids \cite{Bergstra2011,Bergstra2012}. As to the non-identification challenge, the optimization algorithm has been refined significantly in the past years to the extent that it converges to the simple first order stochastic gradient descent with momentum \cite{Kingma2014} and specific initialization methods \cite{Glorot2010,He2015}. As to the local irregularity, robust training methods and monotonicity constraints can be used. To formally measure local irregularity, researchers evaluated model performance on adversarial examples \cite{Goodfellow2015,Kurakin2016,Kurakin2017}. To defend against the adversarial attacks, researchers designed adversarial training methods by incorporating the adversarial attacks into the training process \cite{Kurakin2016}, defensive knowledge distillation \cite{Papernot2016_2}, mini-max robust training \cite{Madry2017}, and even simple gradient regularization \cite{Ross2018}. This study can only open up the discussion about why new methods are necessary for reliable information in DNN-based choice analysis, but a definitive answer needs significant future efforts. 

This study interprets DNN-based choice models by focusing on only the economic information, but economic information is not the only valuable information researchers can obtain from DNNs. Our method of computing the input gradients is the same as the gradient-based methods, which are often referred to under different names such as sensitivity analysis, saliency, or attribution maps in computer vision \cite{Simonyan2013,Kim2017,Smilkov2017,Selvaraju2017,Sundararajan2017,Ross2018}, or attention mechanism in natural language processing \cite{YinWenpeng2016} and generic DNN interpretation literature. But besides gradient-based methods, researchers can also use case-based methods, such as activation maximization (AM), to identify meaningful individual observations to interpret DNN models \cite{Erhan2009,Simonyan2013,Montavon2018,Kim2017}. Researchers can also interpret DNN models by mapping the neurons of the hidden layers to the input space \cite{Zeiler2014} or visualizing the activation maps in the last layer \cite{ZhouBolei2016}. With these approaches, it is possible to extract from DNNs more valuable information beyond the economic information. In short, future researchers should explore all these tools to improve the interpretability of DNN-based choice models and apply the method introduced in this study to a massive number of choice analysis settings. Given the power of DNNs and the infinite opportunities in choice modeling, we believe that the interaction between utility-based choice analysis and DNNs for economic interpretation will be a fertile research area in the future.

\section{Acknowledgement}
\noindent
We thank Singapore-MIT Alliance for Research and Technology (SMART) for partially funding this research. We thank Mary Rose Fissinger for her careful proofreading and Jerry Hausman for inspiring Shenhao with an initial idea of this paper in a casual talk.

\section{Contributions of Authors}
\noindent
S.W. and J.Z. conceived of the presented idea; S.W. developed the theory and reviewed previous studies; S.W. derived the analytical proofs. S.W. and Q.W. designed and conducted the experiments; S.W. drafted the manuscripts; Q.W. and J.Z. provided comments; J.Z. supervised this work. All authors discussed the results and contributed to the final manuscript.

\printbibliography

@article{ZhouBolei2014,
   Author = {Zhou, Bolei and Khosla, Aditya and Lapedriza, Agata and Oliva, Aude and Torralba, Antonio},
   Title = {Object detectors emerge in deep scene cnns},
   Journal = {arXiv preprint arXiv:1412.6856},
   Year = {2014} }

@article{ZhangChiyuan2016,
   Author = {Zhang, Chiyuan and Bengio, Samy and Hardt, Moritz and Recht, Benjamin and Vinyals, Oriol},
   Title = {Understanding deep learning requires rethinking generalization},
   Journal = {arXiv preprint arXiv:1611.03530},
   Year = {2016} }

@inproceedings{Zeiler2014,
   Author = {Zeiler, Matthew D and Fergus, Rob},
   Title = {Visualizing and understanding convolutional networks},
   BookTitle = {European conference on computer vision},
   Publisher = {Springer},
   Pages = {818-833},
   Year = {2014} }

@article{XieChi2003,
   Author = {Xie, Chi and Lu, Jinyang and Parkany, Emily},
   Title = {Work travel mode choice modeling with data mining: decision trees and neural networks},
   Journal = {Transportation Research Record: Journal of the Transportation Research Board},
   Number = {1854},
   Pages = {50-61},
   Year = {2003} }

@incollection{Von_Luxburg2011,
   Author = {Von Luxburg, Ulrike and Schölkopf, Bernhard},
   Title = {Statistical learning theory: Models, concepts, and results},
   BookTitle = {Handbook of the History of Logic},
   Publisher = {Elsevier},
   Volume = {10},
   Pages = {651-706},
   Year = {2011} }

@inproceedings{Vincent2008,
   Author = {Vincent, Pascal and Larochelle, Hugo and Bengio, Yoshua and Manzagol, Pierre-Antoine},
   Title = {Extracting and composing robust features with denoising autoencoders},
   BookTitle = {Proceedings of the 25th international conference on Machine learning},
   Publisher = {ACM},
   Pages = {1096-1103},
   Year = {2008} }

@article{Vapnik1999,
   Author = {Vapnik, Vladimir Naumovich},
   Title = {An overview of statistical learning theory},
   Journal = {IEEE transactions on neural networks},
   Volume = {10},
   Number = {5},
   Pages = {988-999},
   Year = {1999} }

@book{Train2009,
   Author = {Train, Kenneth E},
   Title = {Discrete choice methods with simulation},
   Publisher = {Cambridge university press},
   Year = {2009} }

@inproceedings{Szegedy2015,
   Author = {Szegedy, Christian and Liu, Wei and Jia, Yangqing and Sermanet, Pierre and Reed, Scott and Anguelov, Dragomir and Erhan, Dumitru and Vanhoucke, Vincent and Rabinovich, Andrew},
   Title = {Going deeper with convolutions},
   Publisher = {Cvpr},
   Year = {2015} }

@incollection{Small2007td,
   Author = {Small, Kenneth A and Verhoef, Erik T and Lindsey, Robin},
   Title = {Travel Demand},
   BookTitle = {The economics of urban transportation},
   Publisher = {Routledge},
   Volume = {2},
   Year = {2007} }

@incollection{Small1998,
   Author = {Small, Kenneth and Winston, Clifford},
   Title = {The demand for transportation: models and applications},
   BookTitle = {Essays in Transportation Economics and Policy},
   Year = {1998} }

@article{Sekhar2016,
   Author = {Sekhar, Ch Ravi and Madhu, E},
   Title = {Mode Choice Analysis Using Random Forrest Decision Trees},
   Journal = {Transportation Research Procedia},
   Volume = {17},
   Pages = {644-652},
   Year = {2016} }

@inproceedings{Ribeiro2016,
   Author = {Ribeiro, Marco Tulio and Singh, Sameer and Guestrin, Carlos},
   Title = {Why should i trust you?: Explaining the predictions of any classifier},
   BookTitle = {Proceedings of the 22nd ACM SIGKDD International Conference on Knowledge Discovery and Data Mining},
   Publisher = {ACM},
   Pages = {1135-1144},
   Year = {2016} }

@article{Pulugurta2013,
   Author = {Pulugurta, Sarada and Arun, Ashutosh and Errampalli, Madhu},
   Title = {Use of artificial intelligence for mode choice analysis and comparison with traditional multinomial logit model},
   Journal = {Procedia-Social and Behavioral Sciences},
   Volume = {104},
   Pages = {583-592},
   Year = {2013} }

@inproceedings{Paredes2017,
   Author = {Paredes, Miguel and Hemberg, Erik and O'Reilly, Una-May and Zegras, Chris},
   Title = {Machine learning or discrete choice models for car ownership demand estimation and prediction?},
   BookTitle = {Models and Technologies for Intelligent Transportation Systems (MT-ITS), 2017 5th IEEE International Conference on},
   Publisher = {IEEE},
   Pages = {780-785},
   Year = {2017} }

@book{Ortuzar2011,
   Author = {De Dios Ortuzar, Juan and Willumsen, Luis G},
   Title = {Modelling transport},
   Publisher = {John Wiley and Sons},
   Year = {2011} }

@article{Omrani2015,
   Author = {Omrani, Hichem},
   Title = {Predicting travel mode of individuals by machine learning},
   Journal = {Transportation Research Procedia},
   Volume = {10},
   Pages = {840-849},
   Year = {2015} }

@inproceedings{Nguyen2015,
   Author = {Nguyen, Anh and Yosinski, Jason and Clune, Jeff},
   Title = {Deep neural networks are easily fooled: High confidence predictions for unrecognizable images},
   BookTitle = {Proceedings of the IEEE Conference on Computer Vision and Pattern Recognition},
   Pages = {427-436},
   Year = {2015} }

@article{Mullainathan2017,
   Author = {Mullainathan, Sendhil and Spiess, Jann},
   Title = {Machine learning: an applied econometric approach},
   Journal = {Journal of Economic Perspectives},
   Volume = {31},
   Number = {2},
   Pages = {87-106},
   Year = {2017} }

@article{McFadden1974,
   Author = {McFadden, Daniel},
   Title = {Conditional logit analysis of qualitative choice behavior},
   Year = {1974} }

@article{Lipton2016,
   Author = {Lipton, Zachary C},
   Title = {The mythos of model interpretability},
   Journal = {arXiv preprint arXiv:1606.03490},
   Year = {2016} }

@article{LeCun2015,
   Author = {LeCun, Yann and Bengio, Yoshua and Hinton, Geoffrey},
   Title = {Deep learning},
   Journal = {Nature},
   Volume = {521},
   Number = {7553},
   Pages = {436-444},
   Year = {2015} }

@inproceedings{Krizhevsky2012,
   Author = {Krizhevsky, Alex and Sutskever, Ilya and Hinton, Geoffrey E},
   Title = {Imagenet classification with deep convolutional neural networks},
   BookTitle = {Advances in neural information processing systems},
   Pages = {1097-1105},
   Year = {2012} }

@article{Kotsiantis2007,
   Author = {Kotsiantis, Sotiris B and Zaharakis, I and Pintelas, P},
   Title = {Supervised machine learning: A review of classification techniques},
   Journal = {Emerging artificial intelligence applications in computer engineering},
   Volume = {160},
   Pages = {3-24},
   Year = {2007} }

@article{Kingma2014,
   Author = {Kingma, Diederik P and Ba, Jimmy},
   Title = {Adam: A method for stochastic optimization},
   Journal = {arXiv preprint arXiv:1412.6980},
   Year = {2014} }

@article{Karlaftis2011,
   Author = {Karlaftis, Matthew G and Vlahogianni, Eleni I},
   Title = {Statistical methods versus neural networks in transportation research: Differences, similarities and some insights},
   Journal = {Transportation Research Part C: Emerging Technologies},
   Volume = {19},
   Number = {3},
   Pages = {387-399},
   Year = {2011} }

@article{Hornik1989,
   Author = {Hornik, Kurt and Stinchcombe, Maxwell and White, Halbert},
   Title = {Multilayer feedforward networks are universal approximators},
   Journal = {Neural networks},
   Volume = {2},
   Number = {5},
   Pages = {359-366},
   Year = {1989} }

@article{Hinton2015,
   Author = {Hinton, Geoffrey and Vinyals, Oriol and Dean, Jeff},
   Title = {Distilling the knowledge in a neural network},
   Journal = {arXiv preprint arXiv:1503.02531},
   Year = {2015} }

@article{Hinton2012,
   Author = {Hinton, Geoffrey E and Srivastava, Nitish and Krizhevsky, Alex and Sutskever, Ilya and Salakhutdinov, Ruslan R},
   Title = {Improving neural networks by preventing co-adaptation of feature detectors},
   Journal = {arXiv preprint arXiv:1207.0580},
   Year = {2012} }

@article{Helveston2015,
   Author = {Helveston, John Paul and Liu, Yimin and Feit, Elea McDonnell and Fuchs, Erica and Klampfl, Erica and Michalek, Jeremy J},
   Title = {Will subsidies drive electric vehicle adoption? Measuring consumer preferences in the US and China},
   Journal = {Transportation Research Part A: Policy and Practice},
   Volume = {73},
   Pages = {96-112},
   Year = {2015} }

@inproceedings{He2015,
   Author = {He, Kaiming and Zhang, Xiangyu and Ren, Shaoqing and Sun, Jian},
   Title = {Delving deep into rectifiers: Surpassing human-level performance on imagenet classification},
   BookTitle = {Proceedings of the IEEE international conference on computer vision},
   Pages = {1026-1034},
   Year = {2015} }

@article{Hagenauer2017,
   Author = {Hagenauer, Julian and Helbich, Marco},
   Title = {A comparative study of machine learning classifiers for modeling travel mode choice},
   Journal = {Expert Systems with Applications},
   Volume = {78},
   Pages = {273-282},
   Year = {2017} }

@book{Goodfellow2016,
   Author = {Goodfellow, Ian and Bengio, Yoshua and Courville, Aaron and Bengio, Yoshua},
   Title = {Deep learning},
   Publisher = {MIT press Cambridge},
   Volume = {1},
   Year = {2016} }

@inproceedings{Glorot2010,
   Author = {Glorot, Xavier and Bengio, Yoshua},
   Title = {Understanding the difficulty of training deep feedforward neural networks},
   BookTitle = {Proceedings of the thirteenth international conference on artificial intelligence and statistics},
   Pages = {249-256},
   Year = {2010} }

@book{Geron2017,
   Author = {Géron, Aurélien},
   Title = {Hands-on machine learning with Scikit-Learn and TensorFlow: concepts, tools, and techniques to build intelligent systems},
   Publisher = {" O'Reilly Media, Inc."},
   Year = {2017} }

@article{Fernandez2014,
   Author = {Fernández-Delgado, Manuel and Cernadas, Eva and Barro, Senén and Amorim, Dinani},
   Title = {Do we need hundreds of classifiers to solve real world classification problems},
   Journal = {Journal of Machine Learning Research},
   Volume = {15},
   Number = {1},
   Pages = {3133-3181},
   Year = {2014} }

@techreport{Cohen2016,
   Author = {Cohen, Jonathan D and Ericson, Keith Marzilli and Laibson, David and White, John Myles},
   Title = {Measuring time preferences},
   Institution = {National Bureau of Economic Research},
   Year = {2016} }

@article{Celikoglu2006,
   Author = {Celikoglu, Hilmi Berk},
   Title = {Application of radial basis function and generalized regression neural networks in non-linear utility function specification for travel mode choice modelling},
   Journal = {Mathematical and Computer Modelling},
   Volume = {44},
   Number = {7},
   Pages = {640-658},
   Year = {2006} }

@book{Boyd2004,
   Author = {Boyd, Stephen and Vandenberghe, Lieven},
   Title = {Convex optimization},
   Publisher = {Cambridge university press},
   Year = {2004} }

@article{Boshi_Velez2017,
   Author = {Doshi-Velez, Finale and Kim, Been},
   Title = {Towards a rigorous science of interpretable machine learning},
   Year = {2017} }

@book{Bishop2006,
   Author = {Bishop, Christopher M},
   Title = {Pattern recognition and machine learning},
   Publisher = {springer},
   Year = {2006} }

@book{Ben_Akiva2014,
   Author = {Ben-Akiva, Moshe and Bierlaire, Michel and McFadden, Daniel and Walker, Joan},
   Title = {Discrete Choice Analysis},
   Year = {2014} }

@article{Ben_Akiva1996,
   Author = {Ben-Akiva, Moshe and Bowman, John L and Gopinath, Dinesh},
   Title = {Travel demand model system for the information era},
   Journal = {Transportation},
   Volume = {23},
   Number = {3},
   Pages = {241-266},
   Year = {1996} }

@book{Ben_Akiva1985,
   Author = {Ben-Akiva, Moshe E and Lerman, Steven R},
   Title = {Discrete choice analysis: theory and application to travel demand},
   Publisher = {MIT press},
   Volume = {9},
   Year = {1985} }

@article{Baker2016,
   Author = {Baker, Bowen and Gupta, Otkrist and Naik, Nikhil and Raskar, Ramesh},
   Title = {Designing neural network architectures using reinforcement learning},
   Journal = {arXiv preprint arXiv:1611.02167},
      Year = {2016} }

@article{Bergstra2012,
   Author = {Bergstra, James and Bengio, Yoshua},
   Title = {Random search for hyper-parameter optimization},
   Journal = {Journal of Machine Learning Research},
   Volume = {13},
   Number = {Feb},
   Pages = {281-305},
      Year = {2012} }

@inproceedings{Bergstra2011,
   Author = {Bergstra, James S and Bardenet, Rémi and Bengio, Yoshua and Kégl, Balázs},
   Title = {Algorithms for hyper-parameter optimization},
   BookTitle = {Advances in neural information processing systems},
   Pages = {2546-2554},
      Year = {2011} }

@article{Zoph2016,
   Author = {Zoph, Barret and Le, Quoc V},
   Title = {Neural architecture search with reinforcement learning},
   Journal = {arXiv preprint arXiv:1611.01578},
      Year = {2016} }

@article{Zoph2017,
   Author = {Zoph, Barret and Vasudevan, Vijay and Shlens, Jonathon and Le, Quoc V},
   Title = {Learning transferable architectures for scalable image recognition},
   Journal = {arXiv preprint arXiv:1707.07012},
   Volume = {2},
   Number = {6},
      Year = {2017} }

@article{Falkner2018,
   Author = {Falkner, Stefan and Klein, Aaron and Hutter, Frank},
   Title = {BOHB: Robust and efficient hyperparameter optimization at scale},
   Journal = {arXiv preprint arXiv:1807.01774},
      Year = {2018} }

@article{LiLisha2017,
   Author = {Li, Lisha and Jamieson, Kevin and DeSalvo, Giulia and Rostamizadeh, Afshin and Talwalkar, Ameet},
   Title = {Hyperband: A novel bandit-based approach to hyperparameter optimization},
   Journal = {The Journal of Machine Learning Research},
   Volume = {18},
   Number = {1},
   Pages = {6765-6816},
      Year = {2017} }

@inproceedings{Snoek2012,
   Author = {Snoek, Jasper and Larochelle, Hugo and Adams, Ryan P},
   Title = {Practical bayesian optimization of machine learning algorithms},
   BookTitle = {Advances in neural information processing systems},
   Pages = {2951-2959},
      Year = {2012} }

@inproceedings{Snoek2015,
   Author = {Snoek, Jasper and Rippel, Oren and Swersky, Kevin and Kiros, Ryan and Satish, Nadathur and Sundaram, Narayanan and Patwary, Mostofa and Prabhat, Mr and Adams, Ryan},
   Title = {Scalable bayesian optimization using deep neural networks},
   BookTitle = {International Conference on Machine Learning},
   Pages = {2171-2180},
      Year = {2015} }

@inproceedings{HeKaiming2016, 
   Author = {He, Kaiming and Zhang, Xiangyu and Ren, Shaoqing and Sun, Jian},
   Title = {Deep residual learning for image recognition},
   BookTitle = {Proceedings of the IEEE conference on computer vision and pattern recognition},
   Pages = {770-778},
      Year = {2016} }

@article{Guan2018, 
   Author = {Annaswamy, Anuradha M and Guan, Yue and Tseng, H Eric and Zhou, Hao and Phan, Thao and Yanakiev, Diana},
   Title = {Transactive Control in Smart Cities},
   Journal = {Proceedings of the IEEE},
   Volume = {106},
   Number = {4},
   Pages = {518-537},
      Year = {2018} }

@article{Cantarella2005,
   Author = {Cantarella, Giulio Erberto and de Luca, Stefano},
   Title = {Multilayer feedforward networks for transportation mode choice analysis: An analysis and a comparison with random utility models},
   Journal = {Transportation Research Part C: Emerging Technologies},
   Volume = {13},
   Number = {2},
   Pages = {121-155},
      Year = {2005} }

@article{XiaoGuangnian2016,
   Author = {Xiao, Guangnian and Juan, Zhicai and Zhang, Chunqin},
   Title = {Detecting trip purposes from smartphone-based travel surveys with artificial neural networks and particle swarm optimization},
   Journal = {Transportation Research Part C: Emerging Technologies},
   Volume = {71},
   Pages = {447-463},
      Year = {2016} }

@article{LiuLijuan2017,
   Author = {Liu, Lijuan and Chen, Rung-Ching},
   Title = {A novel passenger flow prediction model using deep learning methods},
   Journal = {Transportation Research Part C: Emerging Technologies},
   Volume = {84},
   Pages = {74-91},
      Year = {2017} }

@article{Polson2017,
   Author = {Polson, Nicholas G and Sokolov, Vadim O},
   Title = {Deep learning for short-term traffic flow prediction},
   Journal = {Transportation Research Part C: Emerging Technologies},
   Volume = {79},
   Pages = {1-17},
      Year = {2017} }

@article{HuangXiuling2018,
   Author = {Huang, Xiuling and Sun, Jie and Sun, Jian},
   Title = {A car-following model considering asymmetric driving behavior based on long short-term memory neural networks},
   Journal = {Transportation Research Part C: Emerging Technologies},
   Volume = {95},
   Pages = {346-362},
      Year = {2018} }

@article{WuYuankai2018,
   Author = {Wu, Yuankai and Tan, Huachun and Qin, Lingqiao and Ran, Bin and Jiang, Zhuxi},
   Title = {A hybrid deep learning based traffic flow prediction method and its understanding},
   Journal = {Transportation Research Part C: Emerging Technologies},
   Volume = {90},
   Pages = {166-180},
      Year = {2018} }

@article{ZhangZhenhua2018,
   Author = {Zhang, Zhenhua and He, Qing and Gao, Jing and Ni, Ming},
   Title = {A deep learning approach for detecting traffic accidents from social media data},
   Journal = {Transportation research part C: emerging technologies},
   Volume = {86},
   Pages = {580-596},
      Year = {2018} }

@article{Cranenburgh2019,
   Author = {van Cranenburgh, Sander and Alwosheel, Ahmad},
   Title = {An artificial neural network based approach to investigate travellers’ decision rules},
   Journal = {Transportation Research Part C: Emerging Technologies},
   Volume = {98},
   Pages = {152-166},
      Year = {2019} }

@article{Bentz2000,
   Author = {Bentz, Yves and Merunka, Dwight},
   Title = {Neural networks and the multinomial logit for brand choice modelling: a hybrid approach},
   Journal = {Journal of Forecasting},
   Volume = {19},
   Number = {3},
   Pages = {177-200},
      Year = {2000} }

@article{Montavon2018, 
   Author = {Montavon, Gregoire and Samek, Wojciech and Muller, Klaus-Robert},
   Title = {Methods for interpreting and understanding deep neural networks},
   Journal = {Digital Signal Processing},
   Volume = {73},
   Pages = {1-15},
      Year = {2018}}

@book{Anthony2009,
   Author = {Anthony, Martin and Bartlett, Peter L},
   Title = {Neural network learning: Theoretical foundations},
   Publisher = {cambridge university press},
      Year = {2009}}

@book{Vershynin2018,
   Author = {Vershynin, Roman},
   Title = {High-dimensional probability: An introduction with applications in data science},
   Publisher = {Cambridge University Press},
   Volume = {47},
      Year = {2018}}

@book{Wainwright2019,
   Author = {Wainwright, Martin J},
   Title = {High-dimensional statistics: A non-asymptotic viewpoint},
   Publisher = {Cambridge University Press},
   Volume = {48},
      Year = {2019}}

@article{Bartlett2002,
   Author = {Bartlett, Peter L and Mendelson, Shahar},
   Title = {Rademacher and Gaussian complexities: Risk bounds and structural results},
   Journal = {Journal of Machine Learning Research},
   Volume = {3},
   Number = {Nov},
   Pages = {463-482},
      Year = {2002}}

@article{Bartlett2006,
   Author = {Bartlett, Peter L and Jordan, Michael I and McAuliffe, Jon D},
   Title = {Convexity, classification, and risk bounds},
   Journal = {Journal of the American Statistical Association},
   Volume = {101},
   Number = {473},
   Pages = {138-156},
      Year = {2006} }

@article{Bartlett2017,
   Author = {Bartlett, Peter L and Harvey, Nick and Liaw, Chris and Mehrabian, Abbas},
   Title = {Nearly-tight VC-dimension and pseudodimension bounds for piecewise linear neural networks},
   Journal = {arXiv preprint arXiv:1703.02930},
      Year = {2017} }

@article{Golowich2017,
   Author = {Golowich, Noah and Rakhlin, Alexander and Shamir, Ohad},
   Title = {Size-independent sample complexity of neural networks},
   Journal = {arXiv preprint arXiv:1712.06541},
      Year = {2017} }

@article{Haussler1995,
   Author = {Haussler, David and Long, Philip M},
   Title = {A generalization of Sauer's lemma},
   Journal = {Journal of Combinatorial Theory, Series A},
   Volume = {71},
   Number = {2},
   Pages = {219-240},
      Year = {1995}}

@inproceedings{Neyshabur2015,
   Author = {Neyshabur, Behnam and Tomioka, Ryota and Srebro, Nathan},
   Title = {Norm-based capacity control in neural networks},
   BookTitle = {Conference on Learning Theory},
   Pages = {1376-1401},
      Year = {2015} }

@article{Seo2017,
   Author = {Seo, Toru and Kusakabe, Takahiko and Gotoh, Hiroto and Asakura, Yasuo},
   Title = {Interactive online machine learning approach for activity-travel survey},
   Journal = {Transportation Research Part B: Methodological},
      Year = {2017}}

@article{Mozolin2000,
   Author = {Mozolin, Mikhail and Thill, J-C and Usery, E Lynn},
   Title = {Trip distribution forecasting with multilayer perceptron neural networks: A critical evaluation},
   Journal = {Transportation Research Part B: Methodological},
   Volume = {34},
   Number = {1},
   Pages = {53-73},
      Year = {2000}}

@article{Duan2016,
   Author = {Duan, Yanjie and Lv, Yisheng and Liu, Yu-Liang and Wang, Fei-Yue},
   Title = {An efficient realization of deep learning for traffic data imputation},
   Journal = {Transportation research part C: emerging technologies},
   Volume = {72},
   Pages = {168-181},
      Year = {2016}}

@article{Cybenko1989,
   Author = {Cybenko, George},
   Title = {Approximation by superpositions of a sigmoidal function},
   Journal = {Mathematics of control, signals and systems},
   Volume = {2},
   Number = {4},
   Pages = {303-314},
      Year = {1989} }

@article{Hornik1991,
   Author = {Hornik, Kurt},
   Title = {Approximation capabilities of multilayer feedforward networks},
   Journal = {Neural networks},
   Volume = {4},
   Number = {2},
   Pages = {251-257},
      Year = {1991}}

@article{Poggio2017, 
   Author = {Poggio, Tomaso and Mhaskar, Hrushikesh and Rosasco, Lorenzo and Miranda, Brando and Liao, Qianli},
   Title = {Why and when can deep-but not shallow-networks avoid the curse of dimensionality: a review},
   Journal = {International Journal of Automation and Computing},
   Volume = {14},
   Number = {5},
   Pages = {503-519},
      Year = {2017} }

@article{Rolnick2017, 
   Author = {Rolnick, David and Tegmark, Max},
   Title = {The power of deeper networks for expressing natural functions},
   Journal = {arXiv preprint arXiv:1705.05502},
      Year = {2017} }

@incollection{Bousquet2004,
   Author = {Bousquet, Olivier and Boucheron, Stéphane and Lugosi, Gábor},
   Title = {Introduction to statistical learning theory},
   BookTitle = {Advanced lectures on machine learning},
   Publisher = {Springer},
   Pages = {169-207},
      Year = {2004} }

@book{Ledoux2013,
   Author = {Ledoux, Michel and Talagrand, Michel},
   Title = {Probability in Banach Spaces: isoperimetry and processes},
   Publisher = {Springer Science & Business Media},
      Year = {2013} }

@article{Baehrens2010, 
   Author = {Baehrens, David and Schroeter, Timon and Harmeling, Stefan and Kawanabe, Motoaki and Hansen, Katja and MÃžller, Klaus-Robert},
   Title = {How to explain individual classification decisions},
   Journal = {Journal of Machine Learning Research},
   Volume = {11},
   Number = {Jun},
   Pages = {1803-1831},
      Year = {2010} }

@article{Goodfellow2015,
   Author = {Goodfellow, Ian J and Shlens, Jonathon and Szegedy, Christian},
   Title = {Explaining and harnessing adversarial examples},
   Journal = {arXiv preprint arXiv:1412.6572},
      Year = {2015} }

@article{Papernot2016_2, 
   Author = {Papernot, Nicolas and McDaniel, Patrick and Goodfellow, Ian},
   Title = {Transferability in machine learning: from phenomena to black-box attacks using adversarial samples},
   Journal = {arXiv preprint arXiv:1605.07277},
      Year = {2016}}

@article{Kurakin2017,
   Author = {Kurakin, Alexey and Goodfellow, Ian and Bengio, Samy},
   Title = {Adversarial examples in the physical world},
   Journal = {arXiv preprint arXiv:1607.02533},
      Year = {2017} }

@article{Kurakin2016,
   Author = {Kurakin, Alexey and Goodfellow, Ian and Bengio, Samy},
   Title = {Adversarial machine learning at scale},
   Journal = {arXiv preprint arXiv:1611.01236},
      Year = {2016} }

@article{Madry2017,
   Author = {Madry, Aleksander and Makelov, Aleksandar and Schmidt, Ludwig and Tsipras, Dimitris and Vladu, Adrian},
   Title = {Towards deep learning models resistant to adversarial attacks},
   Journal = {arXiv preprint arXiv:1706.06083},
      Year = {2017}}

@article{Ross2017,
   Author = {Ross, Andrew Slavin and Hughes, Michael C and Doshi-Velez, Finale},
   Title = {Right for the right reasons: Training differentiable models by constraining their explanations},
   Journal = {arXiv preprint arXiv:1703.03717},
      Year = {2017}}

@inproceedings{Ross2018,
   Author = {Ross, Andrew Slavin and Doshi-Velez, Finale},
   Title = {Improving the adversarial robustness and interpretability of deep neural networks by regularizing their input gradients},
   BookTitle = {Thirty-second AAAI conference on artificial intelligence},
      Year = {2018}}

@article{Szegedy2014,
   Author = {Szegedy, Christian and Zaremba, Wojciech and Sutskever, Ilya and Bruna, Joan and Erhan, Dumitru and Goodfellow, Ian and Fergus, Rob},
   Title = {Intriguing properties of neural networks},
   Journal = {arXiv preprint arXiv:1312.6199},
      Year = {2014}}

@article{Nijkamp1996,
   Author = {Nijkamp, Peter and Reggiani, Aura and Tritapepe, Tommaso},
   Title = {Modelling inter-urban transport flows in Italy: A comparison between neural network analysis and logit analysis},
   Journal = {Transportation Research Part C: Emerging Technologies},
   Volume = {4},
   Number = {6},
   Pages = {323-338},
      Year = {1996} }

@article{Rao1998,
   Author = {Rao, PV Subba and Sikdar, PK and Rao, KV Krishna and Dhingra, SL},
   Title = {Another insight into artificial neural networks through behavioural analysis of access mode choice},
   Journal = {Computers, environment and urban systems},
   Volume = {22},
   Number = {5},
   Pages = {485-496},
      Year = {1998} }

@article{Kaewwichian2019,
   Author = {Kaewwichian, Patiphan and Tanwanichkul, Ladda and Pitaksringkarn, Jumrus},
   Title = {Car Ownership Demand Modeling Using Machine Learning: Decision Trees and Neural Networks.},
   Journal = {International Journal of Geomate},
   Volume = {17},
   Number = {62},
   Pages = {219-230},
      Year = {2019}}

@inproceedings{Kim2017,
   Author = {Kim, Been and Doshi-Velez, Finale},
   Title = {Interpretable Machine Learning (ICML Tutorials)},
   BookTitle = {International Conference of Machine Learning},
   Address= {Sydney },
      Year = {2017} }

@article{Aamodt1994,
   Author = {Aamodt, Agnar and Plaza, Enric},
   Title = {Case-based reasoning: Foundational issues, methodological variations, and system approaches},
   Journal = {AI communications},
   Volume = {7},
   Number = {1},
   Pages = {39-59},
      Year = {1994}}

@article{Erhan2009,
   Author = {Erhan, Dumitru and Bengio, Yoshua and Courville, Aaron and Vincent, Pascal},
   Title = {Visualizing higher-layer features of a deep network},
   Journal = {University of Montreal},
   Volume = {1341},
   Number = {3},
   Pages = {1},
      Year = {2009} }

@inproceedings{Selvaraju2017,
   Author = {Selvaraju, Ramprasaath R and Cogswell, Michael and Das, Abhishek and Vedantam, Ramakrishna and Parikh, Devi and Batra, Dhruv},
   Title = {Grad-cam: Visual explanations from deep networks via gradient-based localization},
   BookTitle = {Proceedings of the IEEE International Conference on Computer Vision},
   Pages = {618-626},
      Year = {2017} }

@article{Simonyan2013,
   Author = {Simonyan, Karen and Vedaldi, Andrea and Zisserman, Andrew},
   Title = {Deep inside convolutional networks: Visualising image classification models and saliency maps},
   Journal = {arXiv preprint arXiv:1312.6034},
      Year = {2013} }

@article{Smilkov2017,
   Author = {Smilkov, Daniel and Thorat, Nikhil and Kim, Been and Viégas, Fernanda and Wattenberg, Martin},
   Title = {Smoothgrad: removing noise by adding noise},
   Journal = {arXiv preprint arXiv:1706.03825},
      Year = {2017} }

@inproceedings{Sundararajan2017,
   Author = {Sundararajan, Mukund and Taly, Ankur and Yan, Qiqi},
   Title = {Axiomatic attribution for deep networks},
   BookTitle = {Proceedings of the 34th International Conference on Machine Learning-Volume 70},
   Publisher = {JMLR. org},
   Pages = {3319-3328},
      Year = {2017} }

@article{Bishop1995,
   Author = {Bishop, Chris M},
   Title = {Training with noise is equivalent to Tikhonov regularization},
   Journal = {Neural computation},
   Volume = {7},
   Number = {1},
   Pages = {108-116},
      Year = {1995} }

@inproceedings{ZhouBolei2016,
   Author = {Zhou, Bolei and Khosla, Aditya and Lapedriza, Agata and Oliva, Aude and Torralba, Antonio},
   Title = {Learning deep features for discriminative localization},
   BookTitle = {Computer Vision and Pattern Recognition (CVPR), 2016 IEEE Conference on},
   Publisher = {IEEE},
   Pages = {2921-2929},
      Year = {2016} }

@inproceedings{Dauphin2014,
   Author = {Dauphin, Yann N and Pascanu, Razvan and Gulcehre, Caglar and Cho, Kyunghyun and Ganguli, Surya and Bengio, Yoshua},
   Title = {Identifying and attacking the saddle point problem in high-dimensional non-convex optimization},
   BookTitle = {Advances in neural information processing systems},
   Pages = {2933-2941},
      Year = {2014} }

@inproceedings{Choromanska2015, 
   Author = {Choromanska, Anna and Henaff, Mikael and Mathieu, Michael and Arous, Gérard Ben and LeCun, Yann},
   Title = {The loss surfaces of multilayer networks},
   BookTitle = {Artificial Intelligence and Statistics},
   Pages = {192-204},
      Year = {2015} }

@article{YinWenpeng2016, 
   Author = {Yin, Wenpeng and Schütze, Hinrich and Xiang, Bing and Zhou, Bowen},
   Title = {Abcnn: Attention-based convolutional neural network for modeling sentence pairs},
   Journal = {Transactions of the Association for Computational Linguistics},
   Volume = {4},
   Pages = {259-272},
      Year = {2016} }

@article{Ho2016,
   Author = {Ho, Chinh Q and Mulley, Corinne and Shiftan, Yoram and Hensher, David A},
   Title = {Vehicle value of travel time savings: Evidence from a group-based modelling approach},
   Journal = {Transportation Research Part A: Policy and Practice},
   Volume = {88},
   Pages = {134-150},
      Year = {2016} }

@incollection{Zamparini2016,
   Author = {Zamparini, Luca and Reggiani, Aura},
   Title = {The value of travel time in passenger and freight transport: an overview},
   BookTitle = {Policy analysis of transport networks},
   Publisher = {Routledge},
   Pages = {161-178},
      Year = {2016} }

@article{WuXin2018,
   author = {Wu, Xin and Guo, Jifu and Xian, Kai and Zhou, Xuesong},
   title = {Hierarchical travel demand estimation using multiple data sources: A forward and backward propagation algorithmic framework on a layered computational graph},
   journal = {Transportation Research Part C: Emerging Technologies},
   volume = {96},
   pages = {321-346},
   ISSN = {0968-090X},
   year = {2018},
   type = {Journal Article}
}

@article{Hansen1990,
   author = {Hansen, Lars Kai and Salamon, Peter},
   title = {Neural network ensembles},
   journal = {IEEE transactions on pattern analysis and machine intelligence},
   volume = {12},
   number = {10},
   pages = {993-1001},
   ISSN = {0162-8828},
   year = {1990},
   type = {Journal Article}
}

@inproceedings{Krogh1995,
   author = {Krogh, Anders and Vedelsby, Jesper},
   title = {Neural network ensembles, cross validation, and active learning},
   booktitle = {Advances in neural information processing systems},
   pages = {231-238},
   year = {1995},
   type = {Conference Proceedings}
}

@article{Tsai2008,
   author = {Tsai, Chih-Fong and Wu, Jhen-Wei},
   title = {Using neural network ensembles for bankruptcy prediction and credit scoring},
   journal = {Expert systems with applications},
   volume = {34},
   number = {4},
   pages = {2639-2649},
   ISSN = {0957-4174},
   year = {2008},
   type = {Journal Article}
}

@article{Irvine2020,
   author = {Irvine, Naomi and Nugent, Chris and Zhang, Shuai and Wang, Hui and NG, Wing WY},
   title = {Neural Network Ensembles for Sensor-Based Human Activity Recognition Within Smart Environments},
   journal = {Sensors},
   volume = {20},
   number = {1},
   pages = {216},
   year = {2020},
   type = {Journal Article}
}

@article{Hillel2018,
   author = {Hillel, Tim and Elshafie, Mohammed ZEB and Jin, Ying},
   title = {Recreating passenger mode choice-sets for transport simulation: A case study of London, UK},
   journal = {Proceedings of the Institution of Civil Engineers-Smart Infrastructure and Construction},
   volume = {171},
   number = {1},
   pages = {29-42},
   ISSN = {2397-8759},
   year = {2018},
   type = {Journal Article}
}

@article{Borysov2019,
   author = {Borysov, Stanislav S and Rich, Jeppe and Pereira, Francisco C},
   title = {How to generate micro-agents? A deep generative modeling approach to population synthesis},
   journal = {Transportation Research Part C: Emerging Technologies},
   volume = {106},
   pages = {73-97},
   ISSN = {0968-090X},
   year = {2019},
   type = {Journal Article}
}

@article{ZhangJunbo2018,
   author = {Zhang, Junbo and Zheng, Yu and Qi, Dekang and Li, Ruiyuan and Yi, Xiuwen and Li, Tianrui},
   title = {Predicting citywide crowd flows using deep spatio-temporal residual networks},
   journal = {Artificial Intelligence},
   volume = {259},
   pages = {147-166},
   ISSN = {0004-3702},
   year = {2018},
   type = {Journal Article}
}

@article{DoLoan2019,
   author = {Do, Loan NN and Vu, Hai L and Vo, Bao Q and Liu, Zhiyuan and Phung, Dinh},
   title = {An effective spatial-temporal attention based neural network for traffic flow prediction},
   journal = {Transportation research part C: emerging technologies},
   volume = {108},
   pages = {12-28},
   ISSN = {0968-090X},
   year = {2019},
   type = {Journal Article}
}

@article{HaoSiyu2019,
   author = {Hao, Siyu and Lee, Der-Horng and Zhao, De},
   title = {Sequence to sequence learning with attention mechanism for short-term passenger flow prediction in large-scale metro system},
   journal = {Transportation Research Part C: Emerging Technologies},
   volume = {107},
   pages = {287-300},
   ISSN = {0968-090X},
   year = {2019},
   type = {Journal Article}
}

@article{LeeSeunghyeon2019,
   author = {Lee, Seunghyeon and Xie, Kun and Ngoduy, Dong and Keyvan-Ekbatani, Mehdi},
   title = {An advanced deep learning approach to real-time estimation of lane-based queue lengths at a signalized junction},
   journal = {Transportation research part C: emerging technologies},
   volume = {109},
   pages = {117-136},
   ISSN = {0968-090X},
   year = {2019},
   type = {Journal Article}
}

@article{YangShuguan2019,
   author = {Yang, Shuguan and Ma, Wei and Pi, Xidong and Qian, Sean},
   title = {A deep learning approach to real-time parking occupancy prediction in transportation networks incorporating multiple spatio-temporal data sources},
   journal = {Transportation Research Part C: Emerging Technologies},
   volume = {107},
   pages = {248-265},
   ISSN = {0968-090X},
   year = {2019},
   type = {Journal Article}
}

@article{MaTao2020,
   author = {Ma, Tao and Antoniou, Constantinos and Toledo, Tomer},
   title = {Hybrid machine learning algorithm and statistical time series model for network-wide traffic forecast},
   journal = {Transportation Research Part C: Emerging Technologies},
   volume = {111},
   pages = {352-372},
   ISSN = {0968-090X},
   year = {2020},
   type = {Journal Article}
}

\newpage

\section*{Appendix I: Large Estimation Error of DNNs}
\noindent
\begin{definition} \label{def:excess_error}
Excess error of $\hat{f}$ is defined as
\begin{flalign}
{\E}_S [L(\hat{f}) - L(f^*)] 
\end{flalign}
\end{definition}

\noindent
$L(\hat{f})$ is the population error of the estimator; $L(f^*)$ is the population error of the true model; $L = {\E}_{x,y}[l(y, f(x))]$ and $l(y, f(x))$ is the loss function. Excess error measures to what extent the error of the estimator deviates from the true model, averaged over random sampling $S$. Note that the excess error can be decomposed as following.
\begin{flalign}
{\E}_S [L(\hat{f}) - L(f^*)] = {\E}_S [L(\hat{f}) - L(f_F) + L(f_F) - L(f^*)]
\end{flalign}
\noindent
in which ${\E}_S [L(\hat{f}) - L(f_F)]$ represents the estimation error and ${\E}_S [L(f_F) - L(f^*)]$ represents the approximation error. When the model family $F$ is large enough, the approximation error becomes very small. For the simplicity of our discussion, we assume the approximation error of DNNs equals to zero. As a result, the following discussions use the terms of excess error and estimation error in an interchangable way.

\begin{proposition} \label{prop:vc_bound}
The estimation error of $\hat{f}$ can be bounded by VC dimension
\begin{flalign}
{\E}_S [L_{0/1}(\hat{f}) - L_{0/1}(f^*)] \lesssim O(\frac{v}{N})
\end{flalign}
in which $v$ is the VC dimension of function class $\F$; $N$ is the sample size; $L_{0/1}$ is the binary prediction error.
\end{proposition}

\noindent
\textbf{Proof.} When no misspecification error exists, estimation error can be further decomposed as three terms
\begin{flalign} \label{eq:excess_error_decomposition}
{\E}_S [L(\hat{f}) - L(f^*)] &= {\E}_S [L(\hat{f}) - \hat{L}(\hat{f}) + \hat{L}(\hat{f}) - \hat{L}(f^*) + \hat{L}(f^*) - L(f^*)] \\
	&\leq {\E}_S [L(\hat{f}) - \hat{L}(\hat{f})] \\
	&\leq {\E}_S \ \underset{f \in \F}{\sup} \ [L(f) - \hat{L}(f)] \label{eq:uniform_loss}
\end{flalign}

\noindent
in which $\hat{L}(f) := \frac{1}{N} \sum_i l(y_i,f^*(x_i))$; the first inequality holds because ${\E}_S [\hat{L}(\hat{f}) - \hat{L}(f^*)] \leq 0$ based on the definition of $\hat{f} := \argmin \hat{L}(f)$ and ${\E}_S [\hat{L}(f^*) - L(f^*)] = 0$ based on the law of large numbers; the second inequality holds due to the $\sup$ operator.

Equation \ref{eq:uniform_loss} can be bounded.
\begin{flalign} \label{eq:rad_complexity}
{\E}_S \ \underset{f \in \F}{\sup} \ [L(f) - \hat{L}(f)] &\leq 2 {\E}_{S,\epsilon} \ \underset{f}{\sup} \ \frac{1}{N} \sum_i l(f(x_i), y_i)\epsilon_i
\end{flalign}

\noindent
This proof relies on the technique called symmetrization, as shown in the proof of Theorem 4.10 in \cite{Wainwright2019}. Note that for prediction error, the loss function $l(f(x_i), y_i) = \I \{f(x_i) \neq y_i \} = y_i + (1 - 2y_i) f(x_i)$, as $y_i \in \{0,1 \}$ and $f(x_i) \in \{0,1 \}$. By applying contraction inequality to Equation \ref{eq:rad_complexity},
\begin{flalign}
2 {\E}_{S,\epsilon} \ \underset{f}{\sup} \ \frac{1}{N} \sum_i l(f(x_i), y_i)\epsilon_i &= 2 {\E}_{S,\epsilon} \ \underset{f}{\sup} \ \frac{1}{N} \sum_i (y_i + (1 - 2y_i) f(x_i)) \times \epsilon_i \\
	&\leq 2 {\E}_{S,\epsilon} \ \underset{f}{\sup} \ \frac{1}{N} \sum_i f(x_i)\epsilon_i \\
	&= 2 {\E}_{S} \hat{\R}_N(\F|_S)
\end{flalign}

\noindent
in which the second line uses the contraction inequality \cite{Ledoux2013} and the third uses the definition of Rademacher complexity. Basically the question about the upper bound of estimation error is turned to the question about the complexity of function class of DNN $\F$. There are many ways to derive an upper bound on Rademacher complexity \cite{Bartlett2002}. To obtain the $v/N$ result, Dudley integral and chaining techniques are useful. Let $Z_{f} := \frac{1}{\sqrt{N}} \sum_i \epsilon_i f(x_i)$ and $Z_{g} := \frac{1}{\sqrt{N}} \sum_i \epsilon_i g(x_i)$, in which $f, g \in \F$. Based on Theorem 5.22 Dudley's entropy integral bound in \cite{Wainwright2019},
\begin{flalign}
{\E}_S \ \big[ \underset{f \in \F}{\sup} \ Z_{f} \big] &\leq {\E}_S \ \big[ \underset{f,g \in \F}{\sup} \ Z_{f} - Z_{g} \big] \\
	&\leq 2 {\E}_S \Big[ \underset{f', g' \in \F; \rho_x(f', g')\leq \delta }{\sup} Z_{f'} - Z_{g'} \Big] + 32 \int_{\delta/4}^D \sqrt{\log N_x(u; \F)} du \label{eq: dudley_integral_bound}
\end{flalign}

\noindent
in which $\rho^2_x(f',g') = \frac{1}{N} \sum_{i=1}^{N} (f'(x_i) - g'(x_i))^2$; $f'$ and $g'$ are the components around the $\delta$ distance of one element in the $\delta$ cover of function class $\F$; $D$ is the diameter of the function class $\F$ projected to dataset $S$, defined as $D := \underset{f, g \in \F}{\sup} \rho_x(f, g) \leq 1$; $\delta$ is any positive value in $[0, D]$. Equation \ref{eq: dudley_integral_bound} holds for any $\delta$. The first term in Equation \ref{eq: dudley_integral_bound} measures the local complexity of DNN and the second term measures the error caused by discretization of the function space. The two terms could be bounded separately. For the first term,
\begin{flalign}
{\E}_S \Big[ \underset{f', g' \in \F; \rho_x(f', g')\leq \delta }{\sup} Z_{f'} - Z_{g'} \Big] &= {\E}_S \Big[ \underset{\rho_x(f', g')\leq \delta }{\sup} \frac{1}{\sqrt{N}} \sum_i \epsilon_i (f'(x_i) - g'(x_i)) \Big] \\
	&= \delta {\E}_S ||\epsilon||_2 \\
	&\leq \delta \sqrt{{\E} \sum_i \epsilon_i^2 } \\
	&\leq \delta \sqrt{N}
\end{flalign}

\noindent
in which the second line uses the dual norm; the third line uses the fact that $\epsilon_i$ is a $1$ sub-Gaussian random variable. For the second term in Equation \ref{eq: dudley_integral_bound}, we need to use the Haussler fact \cite{Haussler1995} that
$$ N_x(u; \F) \leq C v(16e)^v(\frac{1}{u})^v $$

\noindent
It implies
\begin{flalign}
32 \int_{\delta/4}^D \sqrt{\log N_x(u; \F)} du &\leq 32 \int_{\delta/4}^D \sqrt{\log \big[ C v(16e)^v(\frac{1}{u})^v \big]} du \\
	&= 32 \int_{\delta/4}^D \sqrt{\log C + \log v + v \log 16e + v \log \frac{1}{u}} \ du \\
	&\leq c_0 \sqrt{v} \int_{\delta/4}^D \sqrt{\log \frac{1}{u}} \ du \\
	&\leq c_0 \sqrt{v} \int_{0}^D \sqrt{\log \frac{1}{u}} \ du \\
	&\leq c'_0 \sqrt{v}
\end{flalign}

\noindent
By plugging in the upper bounds on the two terms back to Equation \ref{eq: dudley_integral_bound} and dividing both side by $\sqrt{N}$, it implies
\begin{flalign}
{\E}_S \ \underset{f \in \F}{\sup} \ \frac{1}{N} \sum_i \epsilon_i f(x_i) &\leq \underset{\delta}{\inf} \Big[\delta + c'_0 \sqrt{\frac{v}{N}} \Big] \\
	&= c'_0 \sqrt{\frac{v}{N}}
\end{flalign}

\noindent
Therefore, the estimation error can be bounded:
\begin{equation}
{\E}_S [L(\hat{f}) - L(f^*)] \lesssim O(\sqrt{\frac{v}{N}})
\end{equation}

\noindent
\textbf{Remarks.} Intuitively, $v/N$ describes the tradeoff between model complexity and sample size. In a typical MNL model, $v$ is of the same scale as the number of parameters and the input dimension $d$; on the contrary, DNN is a much more complex nonlinear model with much larger $v$. As proved by Bartlett (2017) \cite{Bartlett2017}, DNN with $W$ denoting the number of weights and $L$ denoting the depth has VC dimension $O(WL \log(W))$. For instance, when a dataset has $25$ input variables, the VC dimension of a simple DNN with $8$ layers and $100$ neurons as its width is about $320,000$, as opposed to $v = 25$ as the VC dimension of MNL. Theorefore, the theoretical upper bound of DNN on its estimation error is much larger than MNL model. 

Statistical learning theory is a very broad field that can be used to prove the upper bound on the estimation error \cite{Vershynin2018,Wainwright2019}. Proposition \ref{prop:vc_bound} is limited to the binary discrete output, although its extension to multiple classes and continuous output is also possible. The theoretically optimum upper bound on DNN's estimation error is still an ongoing research field. Statisticians have been exploring different methods to bound DNN, and the methods based on empirical process theory and the contraction inequaility could provide the tightest upper bound so far \cite{Golowich2017,Neyshabur2015,Bartlett2002,Ledoux2013}. The proof of tighter bounds based on contraction inequality also relies on the connection between different loss functions, the techniques of margin analysis and surrogate losses \cite{Bartlett2006}. These proofs are beyond the scope of this study.

\section*{Appendix II: Hyperparameter Space}
\begin{table}[ht]
\centering
\caption{Hyperparameter space}
\resizebox{0.8\linewidth}{!}{%
\begin{tabular}{P{0.3\linewidth} | P{0.6\linewidth}}
\toprule
Depth & $[1,2,3,4,5,6,7,8,9,10]$ \\
\hline
Width & $[25, 50, 100, 150, 200]$ \\
\hline
$L_1$ penalty constants & $[0.1, 10^{-2}, 10^{-3}, 10^{-5}, 10^{-10}, 10^{-20}]$ \\
\hline
$L_2$ penalty constants & $[0.1, 10^{-2}, 10^{-3}, 10^{-5}, 10^{-10}, 10^{-20}]$ \\
\hline
Dropout rates & $[0.01, 10^{-5}]$ \\
\bottomrule
\end{tabular}
} 
\label{table:arch_hpo}
\end{table}

\section*{Appendix III: Summary Statistics of Two Datasets}
\noindent
The key statistics of our samples are summarized in Tables \ref{table:summary_sgp} and \ref{table:summary_ld}. 

\begin{table}[ht]
\centering
\caption{Summary statistics of the SGP dataset}
    \begin{tabular}{lrrrrrrr}
        \toprule
        \multicolumn{8}{l}{\textit{Panel 1. Continuous Variables }} \\
        \hline
        &   mean &       std &       min &       25\% &       50\% &       75\% &        max \\
        \midrule
        Walk\_walktime (min)       &  60.504 &  54.875 &   2.0 &  28.00 &  40.0 &  75.00 &  630.0 \\
        Bus\_cost (S\$)            &   2.069 &   1.266 &   0.0 &   1.12 &   1.8 &   2.52 &    7.0 \\
        Bus\_walktime (min)        &  11.964 &  10.782 &   0.0 &   4.20 &   8.0 &  15.00 &   84.0 \\
        Bus\_waittime (min)        &   7.732 &   5.033 &   0.0 &   4.00 &   7.0 &  10.00 &   42.0 \\
        Bus\_ivt (min)             &  25.064 &  18.911 &   0.0 &  10.00 &  21.0 &  31.20 &  168.0 \\
        Ridesharing\_cost (S\$)    &  14.485 &  11.636 &   0.0 &   7.00 &  12.0 &  17.60 &  140.0 \\
        Ridesharing\_waittime (min) &   7.108 &   4.803 &   0.0 &   4.00 &   5.6 &   9.00 &   42.0 \\
        Ridesharing\_ivt (min)     &  18.283 &  13.389 &   0.8 &   9.80 &  15.4 &  23.20 &  147.0 \\
        AV\_cost (S\$)            &  16.076 &  14.598 &   0.0 &   7.70 &  12.1 &  18.70 &  180.0 \\
        AV\_waittime (min)         &   7.249 &   5.675 &   0.0 &   3.00 &   6.0 &   8.00 &   48.0 \\
        AV\_ivt (min)              &  20.115 &  16.989 &   0.6 &   9.00 &  16.2 &  25.20 &  189.0 \\
        Drive\_cost (S\$)          &  10.494 &  10.568 &   0.0 &   3.20 &   7.0 &  16.00 &   70.0 \\
        Drive\_walktime (min)      &   3.968 &   4.176 &   0.0 &   1.40 &   2.8 &   4.80 &   42.0 \\
        Drive\_ivt (min)           &  17.430 &  14.101 &   0.8 &   8.00 &  14.4 &  22.40 &  168.0 \\
        Age (year)                 &  41.349 &  12.478 &  18.0 &  31.00 &  41.0 &  50.00 &   82.0 \\
        Income (K S\$)                  &   9.827 &   5.013 &   0.0 &   7.00 &   9.0 &  13.50 &   20.0 \\
        Education                 &   3.063 &   2.698 &  0.0 &  0.0 &   4.0 &   5.00 &    7.0 \\
        \hline \hline
        \multicolumn{8}{l}{\textit{Panel 2. Discrete Variables (Counts)}} \\
        \hline
        Gender & \multicolumn{7}{l}{5,190 (1: Male); 3,228 (0: Female)} \\
        Employment & \multicolumn{7}{l}{5,064 (1: Employed); 3,354 (0: Unemployed)} \\
        \bottomrule
    \end{tabular}  
\label{table:summary_sgp}
\end{table}

\begin{table}[]
    \centering
    \caption{Summary statistics of the LD dataset}
    \begin{tabular}{lrrrrrrr}
        \toprule
        \multicolumn{8}{l}{\textit{Panel 1. Continuous Variables}} \\
        \hline
         &      mean &       std &     min &       25\% &       50\% &       75\% &        max \\
        \midrule
        Age (Year)       &    39.462 &    19.227 &   5.000 &    25.000 &    38.000 &    52.000 &     99.000 \\
        Distance (Meters)    &  4605 &  4782 &  77 &  1309 &  2814 &  6175 &  40941 \\
        Duration\_walking (h)  & 1.129 &     1.118 &   0.025 &     0.351 &     0.723 &     1.514 &      9.278 \\
        Duration\_cycling (h)  &  0.362 &     0.352 &   0.006 &     0.117 &     0.232 &     0.485 &      3.052 \\
        Duration\_PT\_access (h) & 0.160 &     0.092 &   0.000 &     0.092 &     0.144 &     0.211 &      1.189 \\
        Duration\_PT\_in\_vehicle (h)  &  0.262 &     0.230 &   0.000 &     0.083 &     0.192 &     0.383 &      2.147 \\
        Duration\_pt\_transfer\_total (h) & 0.044 &     0.078 &   0.000 &     0.000 &     0.000 &     0.083 & 0.865 \\
        Duration\_driving (h)        &     0.282 &     0.252 &   0.000 &     0.108 &     0.192 &     0.369 &      2.061 \\
        Cost\_transit (\pounds)       &     1.563 &     1.535 &   0.000 &     0.000 &     1.500 &     2.400 &     13.490 \\
        Cost\_driving\_total (\pounds)&     1.903 &     3.485 &   0.000 &     0.290 &     0.570 &     1.290 &     17.160 \\
        \hline \hline
        \multicolumn{8}{l}{\textit{Panel 2. Discrete Variables (Counts)}} \\
        \hline
        Gender & \multicolumn{7}{l}{42,690 (0: Female); 38,396 (1: Male)} \\
        Driving License & \multicolumn{7}{l}{31,051 (0: No); 50,035 (1: Yes)} \\
        Car Ownership & \multicolumn{7}{l}{23707 (0 Car); 35,229 (1 Car); 22,150 (2 Cars)} \\
        Number of Transfers in PT & \multicolumn{7}{l}{56,668 (0); 19,765 (1); 4,428 (2); 495 (3); 30 (4)} \\
        \bottomrule
    \end{tabular}
\label{table:summary_ld}
\end{table}

\section*{Appendix IV: Formula to Compute Elasticities}
\noindent
This appendix shows the formula used to compute elasticities. There are four types of formula, depending on the type of variable and utility functions. Recall that the utility functions are:

\begin{flalign} \label{eq:util_mnl_1}
& V_{ik} = \beta_{0,k} + \beta_{x,k}^T x_{ik} + \beta_{z,k}^T z_i, \ \ as \ \ k \neq ref \\
& V_{ik} = \beta_{x,k}^T x_{ik}, \ \ as \ \ k = ref 
\end{flalign}

\noindent
For simplicity, the following derivation omits the subscript $i$, uses $x_{k}$ as a scalar to represent an alternative-specific variable, and uses $z$ as a scalar to represent an individual-specific variable.

\begin{enumerate}
    \item Self-elasticity of choice probability $s_k$ with respect to an alternative-specific variable $x_k$. This formula can be found in Train (2009) \cite{Train2009}.
    \begin{flalign}
    \frac{\partial s_k / s_k}{\partial x_k / x_k} & = \frac{\partial s_k}{\partial x_k} \times \frac{x_k}{s_k} \\
        & = \frac{\partial V_k}{\partial x_k} s_k (1 - s_k) \times \frac{x_k}{s_k} \\
        & = \beta_{x,k} x_k (1 - s_k)
    \end{flalign}
    \item Cross-elasticity of choice probability $s_k$ with respect to an alternative-specific variable $x_{k'}$.  This formula can also be found in Train (2009) \cite{Train2009}. Interestingly, the formula does not depend on the alternative $k$, so conditioning on the same $k'$, the cross-elasticities are the same.
    \begin{flalign}
    \frac{\partial s_k / s_k}{\partial x_{k'} / x_{k'}} & = \frac{\partial s_k}{\partial x_{k'}} \times \frac{x_{k'}}{s_k} \\
        & = - \beta_{x,k'} x_{k'} s_{k'}
    \end{flalign}
    \item Elasticity of choice probability $s_k$ ($k = ref$) with respect to individual-specific variable $z_i$. This formula cannot be found in standard textbooks \cite{Ben_Akiva1985,Train2009}, so we show slightly more detailed derivations. 
    \begin{flalign}
    \frac{\partial s_k / s_k}{\partial z / z} & = [\frac{\partial e^{V_k}}{\partial z} \times \frac{1}{\sum_j e^{V_j}} - e^{V_k}(\sum_j e^{V_j})^{-2} \frac{\partial \sum_j e^{V_j}}{\partial z}] \times \frac{z}{s_k} \\
        & = - \frac{e^{V_k}}{(\sum_j e^{V_j})^2} \times \sum_{j \neq k} e^{V_j} \beta_{z,j} \\
        & = - z \sum_{j \neq k} s_j \beta_{z,j}
    \end{flalign}
    \item Elasticity of choice probability $s_k'$ ($k' \neq ref$) with respect to individual-specific variable $z_i$. Here we use $k$ to denote the reference alternative.
    \begin{flalign}
    \frac{\partial s_{k'} / s_{k'}}{\partial z / z} 
        & = [\frac{\partial e^{V_{k'}}}{\partial z} \times \frac{1}{\sum_j e^{V_j}} - e^{V_{k'}}(\sum_j e^{V_j})^{-2} \frac{\partial \sum_j e^{V_j}}{\partial z}] \times \frac{z}{s_{k'}} \\
        & = [\frac{e^{V_{k'}}}{\sum_j e^{V_j}} \beta_{z,k'} - \frac{e^{V_{k'}}}{(\sum_j     e^{V_j})^2} \times \sum_{j \neq k} \frac{\partial e^{V_j}}{\partial x}] \times \frac{z}{s_{k'}} \\
        & = z \beta_{z,k'} - z \sum_{j \neq k} s_j \beta_{z,j}
    \end{flalign}
\end{enumerate}

\newpage
\section*{Appendix V: Further Details of Experimental Results}
\begin{figure}[ht]
\captionsetup[subfigure]{justification=centering}
\centering
\subfloat[One Opt-DNN (8K-SGP; Training)]{\includegraphics[width=0.3\linewidth]{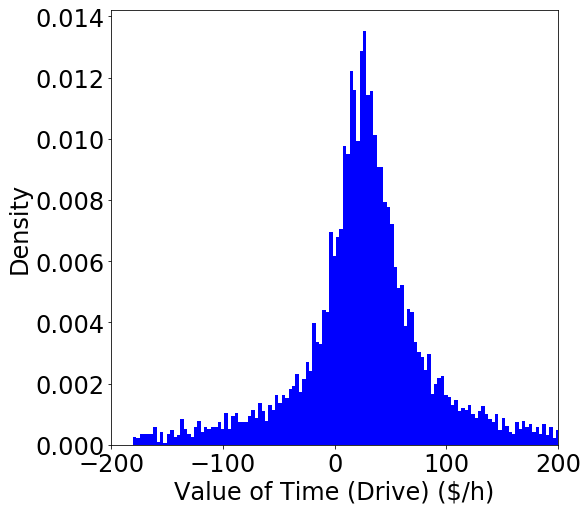}\label{sfig:vot_5l_train}}
\subfloat[One Opt-DNN (80K-LD; Training)]{\includegraphics[width=0.295\linewidth]{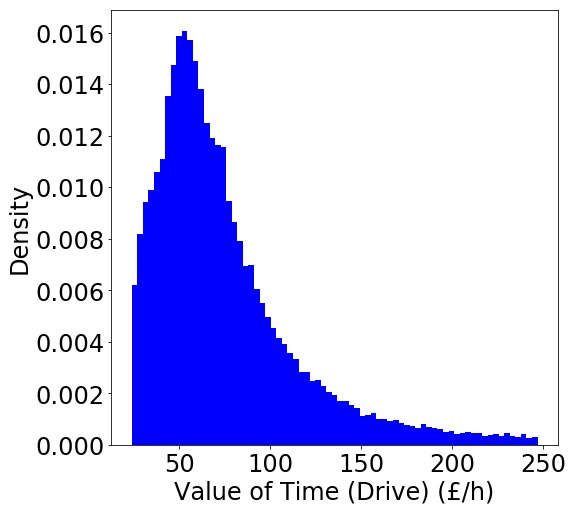}\label{sfig:vot_5l_train}} \\
\caption{Heterogeneous values of time in the training sets; the extremely large and small values are cut-off from this histogram.}
\label{fig:vot_one_training}
\end{figure}

\noindent
Figure \ref{fig:vot_one_training} shows the distributions of the VOT in the training sets of the 8K-SGP and 80K-LD datasets. The distributions in the training sets are very similar to those in the testing sets, suggesting that our previous findings are robust.

\end{document}